\documentclass[12pt,a4paper]{article}
\usepackage{ifthen} % for conditional statements
\newboolean{pdflatex}
\setboolean{pdflatex}{true} % False for eps figures 
\usepackage{subfigure}
\usepackage{afterpage}
\usepackage{overpic}

\newboolean{articletitles}
\setboolean{articletitles}{true} % False removes titles in references

\newboolean{uprightparticles}
\setboolean{uprightparticles}{false} %True for upright particle symbols

\newboolean{inbibliography}
\setboolean{inbibliography}{false} %True once you enter the bibliography

\usepackage{microtype}
\usepackage{lineno}  % for line numbering during review
\usepackage{xspace} % To avoid problems with missing or double spaces after
\usepackage{graphicx}  % to include figures (can also use other packages)
\usepackage{color}
\usepackage{colortbl}
\graphicspath{{./figs/}} % Make Latex search fig subdir for figures

\usepackage{amsmath} % Adds a large collection of math symbols
\usepackage{amssymb}
\usepackage{amsfonts}
\usepackage{upgreek} % Adds in support for greek letters in roman typeset

\newcommand*\patchAmsMathEnvironmentForLineno[1]{%
\expandafter\let\csname old#1\expandafter\endcsname\csname #1\endcsname
\expandafter\let\csname oldend#1\expandafter\endcsname\csname
end#1\endcsname
 \renewenvironment{#1}%
   {\linenomath\csname old#1\endcsname}%
   {\csname oldend#1\endcsname\endlinenomath}%
}
\newcommand*\patchBothAmsMathEnvironmentsForLineno[1]{%
  \patchAmsMathEnvironmentForLineno{#1}%
  \patchAmsMathEnvironmentForLineno{#1*}%
}
\AtBeginDocument{%
\patchBothAmsMathEnvironmentsForLineno{equation}%
\patchBothAmsMathEnvironmentsForLineno{align}%
\patchBothAmsMathEnvironmentsForLineno{flalign}%
\patchBothAmsMathEnvironmentsForLineno{alignat}%
\patchBothAmsMathEnvironmentsForLineno{gather}%
\patchBothAmsMathEnvironmentsForLineno{multline}%
}

\usepackage{hyperref}    % Hyperlinks in references
\usepackage[all]{hypcap} % Internal hyperlinks to floats.

\def\lhcb {\mbox{LHCb}\xspace}

\def\babar  {\mbox{BaBar}\xspace}

\ifthenelse{\boolean{uprightparticles}}%
{

 \def\Ppi         {\ensuremath{\uppi}\xspace}

 \def\Ppsi        {\ensuremath{\uppsi}\xspace}

 \def\PDelta      {\ensuremath{\Delta}\xspace}                 
 \def\PXi      {\ensuremath{\Xi}\xspace}                 
 \def\PLambda      {\ensuremath{\Lambda}\xspace}                 
 \def\PSigma      {\ensuremath{\Sigma}\xspace}                 
 \def\POmega      {\ensuremath{\Omega}\xspace}                 
 \def\PUpsilon      {\ensuremath{\Upsilon}\xspace}                 
 
 \def\PB      {\ensuremath{\mathrm{B}}\xspace}                 
                  
 \def\PD      {\ensuremath{\mathrm{D}}\xspace}

 \def\PJ      {\ensuremath{\mathrm{J}}\xspace}                 
 \def\PK      {\ensuremath{\mathrm{K}}\xspace}

 \def\Pb      {\ensuremath{\mathrm{b}}\xspace}                 
 \def\Pc      {\ensuremath{\mathrm{c}}\xspace}                 
 \def\Pd      {\ensuremath{\mathrm{d}}\xspace}

 \def\Pi      {\ensuremath{\mathrm{i}}\xspace}

 \def\Ps      {\ensuremath{\mathrm{s}}\xspace}                 
                  
 \def\Pu      {\ensuremath{\mathrm{u}}\xspace}

}
{

 \def\Ppi         {\ensuremath{\pi}\xspace}

 \def\Ppsi        {\ensuremath{\psi}\xspace}                 
                  
 \mathchardef\PDelta="7101
 \mathchardef\PXi="7104
 \mathchardef\PLambda="7103
 \mathchardef\PSigma="7106
 \mathchardef\POmega="710A
 \mathchardef\PUpsilon="7107
                  
 \def\PB      {\ensuremath{B}\xspace}                 
                  
 \def\PD      {\ensuremath{D}\xspace}

 \def\PJ      {\ensuremath{J}\xspace}                 
 \def\PK      {\ensuremath{K}\xspace}

 \def\Pb      {\ensuremath{b}\xspace}                 
 \def\Pc      {\ensuremath{c}\xspace}                 
 \def\Pd      {\ensuremath{d}\xspace}

 \def\Pi      {\ensuremath{i}\xspace}

 \def\Ps      {\ensuremath{s}\xspace}                 
                  
 \def\Pu      {\ensuremath{u}\xspace}

}

\def\uquark    {\ensuremath{\Pu}\xspace}
\def\uquarkbar {\ensuremath{\overline \uquark}\xspace}
\def\uubar     {\ensuremath{\uquark\uquarkbar}\xspace}
\def\dquark    {\ensuremath{\Pd}\xspace}
\def\dquarkbar {\ensuremath{\overline \dquark}\xspace}
\def\ddbar     {\ensuremath{\dquark\dquarkbar}\xspace}
\def\squark    {\ensuremath{\Ps}\xspace}
\def\squarkbar {\ensuremath{\overline \squark}\xspace}
\def\ssbar     {\ensuremath{\squark\squarkbar}\xspace}
\def\cquark    {\ensuremath{\Pc}\xspace}

\def\bquark    {\ensuremath{\Pb}\xspace}

\def\pion  {\ensuremath{\Ppi}\xspace}

\def\pip   {\ensuremath{\pion^+}\xspace}
\def\pim   {\ensuremath{\pion^-}\xspace}
\def\pipm  {\ensuremath{\pion^\pm}\xspace}

\def\kaon  {\ensuremath{\PK}\xspace}
  \def\Kbar  {\kern 0.2em\overline{\kern -0.2em \PK}{}\xspace}

\def\Kp    {\ensuremath{\kaon^+}\xspace}
\def\Km    {\ensuremath{\kaon^-}\xspace}
\def\Kpm   {\ensuremath{\kaon^\pm}\xspace}

\def\Kstarz  {\ensuremath{\kaon^{*0}}\xspace}

\def\Kstar   {\ensuremath{\kaon^*}\xspace}

  \def\Dbar    {\kern 0.2em\overline{\kern -0.2em \PD}{}\xspace}
\def\D       {\ensuremath{\PD}\xspace}

\def\Dz      {\ensuremath{\D^0}\xspace}
\def\Dzb     {\ensuremath{\Dbar^0}\xspace}

\def\Dstarp  {\ensuremath{\D^{*+}}\xspace}
\def\Dstarm  {\ensuremath{\D^{*-}}\xspace}

\def\B       {\ensuremath{\PB}\xspace}
\def\Bbar    {\ensuremath{\kern 0.18em\overline{\kern -0.18em \PB}{}}\xspace}

\def\Bu      {\ensuremath{\B^+}\xspace}
\def\Bub     {\ensuremath{\B^-}\xspace}
\def\Bp      {\ensuremath{\Bu}\xspace}
\def\Bm      {\ensuremath{\Bub}\xspace}
\def\Bpm     {\ensuremath{\B^\pm}\xspace}

\def\Bd      {\ensuremath{\B^0}\xspace}

\def\jpsi     {\ensuremath{{\PJ\mskip -3mu/\mskip -2mu\Ppsi\mskip 2mu}}\xspace}

  \def\Y#1S{\ensuremath{\PUpsilon{(#1S)}}\xspace}% no space before {...}!

\def\Lbar {\ensuremath{\kern 0.1em\overline{\kern -0.1em\PLambda}}\xspace}

\def\BF         {{\ensuremath{\cal B}\xspace}}

\def\BR         {\BF}
\newcommand{\decay}[2]{\mbox{\ensuremath{#1\!\to #2}\xspace}}         % {\Pa}{\Pb \Pc}

\def\to                 {\ensuremath{\rightarrow}\xspace}

\def\grpsuthree {\ensuremath{\mathrm{SU}(3)}\xspace}

\def\CP                {\ensuremath{C\!P}\xspace}

\newcommand{\ACP}{\ensuremath{{\cal A}^{\CP}}\xspace}

\def\AT#1     {\ensuremath{A_{\mathrm{T}}^{#1}}\xspace}           % 2

\def\C#1      {\ensuremath{\mathcal{C}_{#1}}\xspace}                       % 9
\def\Cp#1     {\ensuremath{\mathcal{C}_{#1}^{'}}\xspace}                    % 7
\def\Ceff#1   {\ensuremath{\mathcal{C}_{#1}^{\mathrm{(eff)}}}\xspace}        % 9  
\def\Cpeff#1  {\ensuremath{\mathcal{C}_{#1}^{'\mathrm{(eff)}}}\xspace}       % 7
\def\Ope#1    {\ensuremath{\mathcal{O}_{#1}}\xspace}                       % 2
\def\Opep#1   {\ensuremath{\mathcal{O}_{#1}^{'}}\xspace}                    % 7

\newcommand{\ket}[1]{\ensuremath{|#1\rangle}}              % {b}
\newcommand{\tev}{\ifthenelse{\boolean{inbibliography}}{\ensuremath{~T\kern -0.05em eV}\xspace}{\ensuremath{\mathrm{\,Te\kern -0.1em V}}\xspace}}
\newcommand{\gev}{\ensuremath{\mathrm{\,Ge\kern -0.1em V}}\xspace}
\newcommand{\mev}{\ensuremath{\mathrm{\,Me\kern -0.1em V}}\xspace}
\newcommand{\kev}{\ensuremath{\mathrm{\,ke\kern -0.1em V}}\xspace}
\newcommand{\ev}{\ensuremath{\mathrm{\,e\kern -0.1em V}}\xspace}
\newcommand{\gevc}{\ensuremath{{\mathrm{\,Ge\kern -0.1em V\!/}c}}\xspace}
\newcommand{\mevc}{\ensuremath{{\mathrm{\,Me\kern -0.1em V\!/}c}}\xspace}
\newcommand{\gevcc}{\ensuremath{{\mathrm{\,Ge\kern -0.1em V\!/}c^2}}\xspace}
\newcommand{\gevgevcccc}{\ensuremath{{\mathrm{\,Ge\kern -0.1em V^2\!/}c^4}}\xspace}
\newcommand{\mevcc}{\ensuremath{{\mathrm{\,Me\kern -0.1em V\!/}c^2}}\xspace}

\def\mum  {\ensuremath{\,\upmu\rm m}\xspace}

\def\invfb   {\ensuremath{\mbox{\,fb}^{-1}}\xspace}

\newcommand{\chisq}{\ensuremath{\chi^2}\xspace}

\newcommand{\chisqip}{\ensuremath{\chi^2_{\rm IP}}\xspace}

\def\gsim{{~\raise.15em\hbox{$>$}\kern-.85em
          \lower.35em\hbox{$\sim$}~}\xspace}
\def\lsim{{~\raise.15em\hbox{$<$}\kern-.85em
          \lower.35em\hbox{$\sim$}~}\xspace}

\def\pt         {\mbox{$p_{\rm T}$}\xspace}

\def\evtgen     {\mbox{\textsc{EvtGen}}\xspace}

\def\geant      {\mbox{\textsc{Geant4}}\xspace}

\def\photos     {\mbox{\textsc{Photos}}\xspace}

\def\pythia     {\mbox{\textsc{Pythia}}\xspace}

\def\tell1  {TELL1\xspace}
\def\ukl1   {UKL1\xspace}

\newcommand{\eg}{\mbox{\itshape e.g.}\xspace}
\newcommand{\ie}{\mbox{\itshape i.e.}\xspace}

\newcommand{\mycds}{http://cds.cern.ch/record}
\usepackage{cite} % Allows for ranges in citations
\usepackage{mciteplus}

\usepackage{longtable} % only for template; not usually to be used in PAPERs

\begin{document}

\renewcommand{\ACP}{\ensuremath{\mathcal{A}_{C\!P}}}
\newcommand{\A}{\ensuremath{\mathcal{A}}}

%%%%%%%%%%%%%%%%%%%%%%%%%
%%%%% Title     %%%%%%%%%
%%%%%%%%%%%%%%%%%%%%%%%%%
\renewcommand{\thefootnote}{\fnsymbol{footnote}}
\setcounter{footnote}{1}

\newcommand{\subfig}[1]{\subref{fig:#1}}

% %%%%%%% CHOOSE TITLE PAGE--------
%\onecolumn
% \input{title-LHCb-ANA}
%\input{title-LHCb-CONF}
% $Id: title-LHCb-PAPER.tex 35201 2013-05-10 14:33:25Z roldeman $
% ===============================================================================
% Purpose: LHCb-PAPER journal paper title page template
% Author: 
% Created on: 2010-09-25
% ===============================================================================

%%%%%%%%%%%%%%%%%%%%%%%%%
%%%%%  TITLE PAGE  %%%%%%
%%%%%%%%%%%%%%%%%%%%%%%%%
\begin{titlepage}
\pagenumbering{roman}

% Header ---------------------------------------------------
\vspace*{-1.5cm}
\centerline{\large EUROPEAN ORGANIZATION FOR NUCLEAR RESEARCH (CERN)}
\vspace*{1.5cm}
\hspace*{-0.5cm}
\begin{tabular*}{\linewidth}{lc@{\extracolsep{\fill}}r}
\ifthenelse{\boolean{pdflatex}}% Logo format choice
{\vspace*{-2.7cm}\mbox{\!\!\!\includegraphics[width=.14\textwidth]{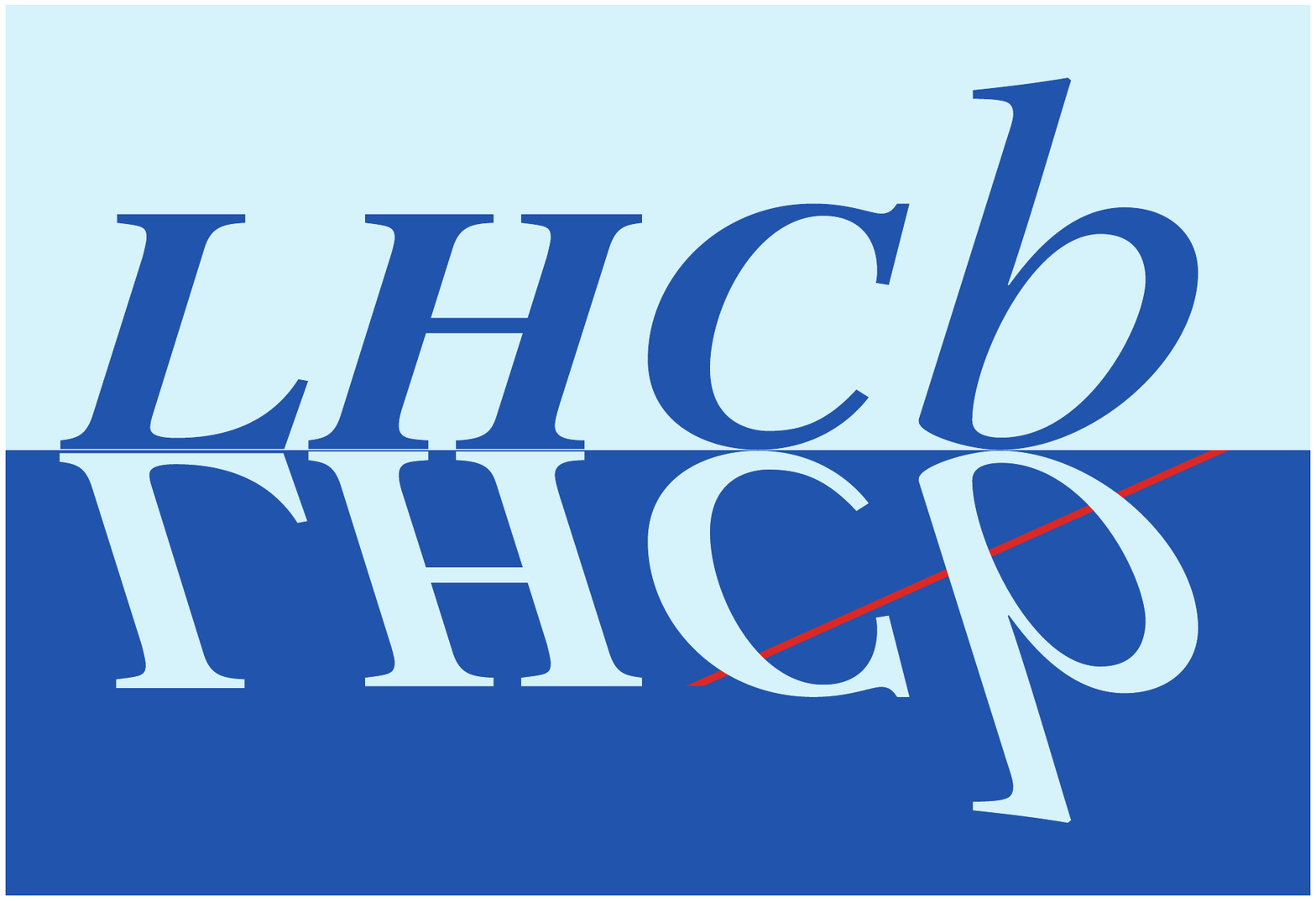}} & &}%
{\vspace*{-1.2cm}\mbox{\!\!\!\includegraphics[width=.12\textwidth]{lhcb-logo.eps}} & &}%
\\
 & & CERN-PH-EP-2013-166 \\  % ID 
 & & LHCb-PAPER-2013-048 \\  % ID 
 & & November 13, 2013 \\ % Date - Can also hardwire e.g.: 23 March 2010
 & & \\
% not in paper \hline
\end{tabular*}

\vspace*{2.0cm}

% Title --------------------------------------------------
{\bf\boldmath\huge
\begin{center}
Measurement of the
charge asymmetry in $B^{\pm}\rightarrow \phi K^{\pm}$ and search for $B^{\pm}\rightarrow \phi \pi^{\pm}$ decays
\end{center}
}

\vspace*{2.0cm}

% Authors -------------------------------------------------
\begin{center}
The LHCb collaboration\footnote{Authors are listed on the following pages.}
\end{center}

\vspace{\fill}

% Abstract -----------------------------------------------
\begin{abstract}
\noindent
 The \CP-violating charge asymmetry in $B^{\pm}\rightarrow \phi K^{\pm}$  decays is measured in   
a sample of $pp$ collisions at 7 TeV centre-of-mass energy, corresponding to an 
integrated luminosity of 1.0 fb$^{-1}$ collected by the LHCb experiment.
The result is $\mathcal{A}_{CP}(B^{\pm}\rightarrow \phi K^{\pm}) =
\rm 0.022\pm 0.021 \pm 0.009$,
where the first uncertainty is statistical and the second systematic. 
In addition, a search for the $B^{\pm}\rightarrow \phi \pi^{\pm}$ decay 
mode is performed, using the $B^{\pm}\rightarrow \phi K^{\pm}$  decay rate for normalization. An upper limit on the branching fraction
$\mathcal{B}(B^{\pm}\rightarrow \phi \pi^{\pm})< 1.5\times 10^{-7}$ is set at 90\% 
confidence level.
\end{abstract}

%\maketitle

\vspace*{2.0cm}

\begin{center}
Submitted to Phys.~Lett.~B 
\end{center}

\vspace{\fill}

{\footnotesize 
\centerline{\copyright~CERN on behalf of the \lhcb collaboration, license \href{http://creativecommons.org/licenses/by/3.0/}{CC-BY-3.0}.}}
\vspace*{2mm}

\end{titlepage}

%%%%%%%%%%%%%%%%%%%%%%%%%%%%%%%%
%%%%%  EOD OF TITLE PAGE  %%%%%%
%%%%%%%%%%%%%%%%%%%%%%%%%%%%%%%%

%  empty page follows the title page ----
\newpage
\setcounter{page}{2}
\mbox{~}
\newpage

% Author List ----------------------------
%  You need to get a new author list!
%\setlength{\bigskipamount}{0mm}
%\vspace{-3cm}
%%%%%%%%%%%%%%%%%%%%%%%%%%%%%%%%%%%%%%%%%%
\centerline{\large\bf LHCb collaboration}
\begin{flushleft}
\small
R.~Aaij$^{40}$, 
B.~Adeva$^{36}$, 
M.~Adinolfi$^{45}$, 
C.~Adrover$^{6}$, 
A.~Affolder$^{51}$, 
Z.~Ajaltouni$^{5}$, 
J.~Albrecht$^{9}$, 
F.~Alessio$^{37}$, 
M.~Alexander$^{50}$, 
S.~Ali$^{40}$
% ...

{\footnotesize \it
$ ^{1}$Centro Brasileiro de Pesquisas F\'{i}sicas (CBPF), Rio de Janeiro, Brazil\\
% ...
$ ^{5}$Clermont Universit\'{e}, Universit\'{e} Blaise Pascal, CNRS/IN2P3, LPC, Clermont-Ferrand, France\\
$ ^{6}$CPPM, Aix-Marseille Universit\'{e}, CNRS/IN2P3, Marseille, France\\
% ...
$ ^{9}$Fakult\"{a}t Physik, Technische Universit\"{a}t Dortmund, Dortmund, Germany\\
% ...
$ ^{36}$Universidad de Santiago de Compostela, Santiago de Compostela, Spain\\
$ ^{37}$European Organization for Nuclear Research (CERN), Geneva, Switzerland\\
% ...
$ ^{40}$Nikhef National Institute for Subatomic Physics, Amsterdam, The Netherlands\\
% ...
$ ^{45}$H.H. Wills Physics Laboratory, University of Bristol, Bristol, United Kingdom\\
% ...
$ ^{50}$School of Physics and Astronomy, University of Glasgow, Glasgow, United Kingdom\\
$ ^{51}$Oliver Lodge Laboratory, University of Liverpool, Liverpool, United Kingdom\\
% ...
}
\end{flushleft}
%%%%%%%%%%%%%%%%%%%%%%%%%%%%%%%%%%%%%%%%%%

% The author list for journal publications is provided by the Membership Committee shortly after 'approval to go to paper' has been given.
% It will be made available on the page
% \verb!http://www.physik.uzh.ch/~strauman/forMemCo/LHCb-PAPER-XXXX-XXX/! .
% The author list should be included already at first circulation, 
% to allow new members of the collaboration to verify whether they have been included correctly.
% Before every new circulation and before submitting, check if the author list has been updated, since occasionally a misspelled name 
% is corrected or associated institutions become full members.
% In case line numbering doesn't work well after including the authorlist, try moving the \verb!\bigskip! after the last author to a separate line.

\cleardoublepage
~
\cleardoublepage

%\twocolumn
% %%%%%%%%%%%%% ---------

\renewcommand{\thefootnote}{\arabic{footnote}}
\setcounter{footnote}{0}

%%%%%%%%%%%%%%%%%%%%%%%%%%%%%%%%
%%%%%  Table of Content   %%%%%%
%%%%%%%%%%%%%%%%%%%%%%%%%%%%%%%%
%%%% Uncomment next 2 lines if desired
%\tableofcontents
%\cleardoublepage

%%%%%%%%%%%%%%%%%%%%%%%%%
%%%%% Main text %%%%%%%%%
%%%%%%%%%%%%%%%%%%%%%%%%%

\pagestyle{plain} % restore page numbers for the main text
\setcounter{page}{1}
\pagenumbering{arabic}

%% Uncomment during review phase. 
%% Comment before a final submission.
%\linenumbers

% You can include short sections directly in the main tex file.
% However, for larger papers it is desirable to split the text into
% several semiautonomous files, which can be revised independently.
% This is especially useful when developing a document in
% collaboration with several people, since then different parts can be
% edited independently.  This type of file organization is shown here.
% 

\section{Introduction}
\label{Introduction}

\begin{figure}[b]
\centering
\includegraphics[width=0.45\textwidth]{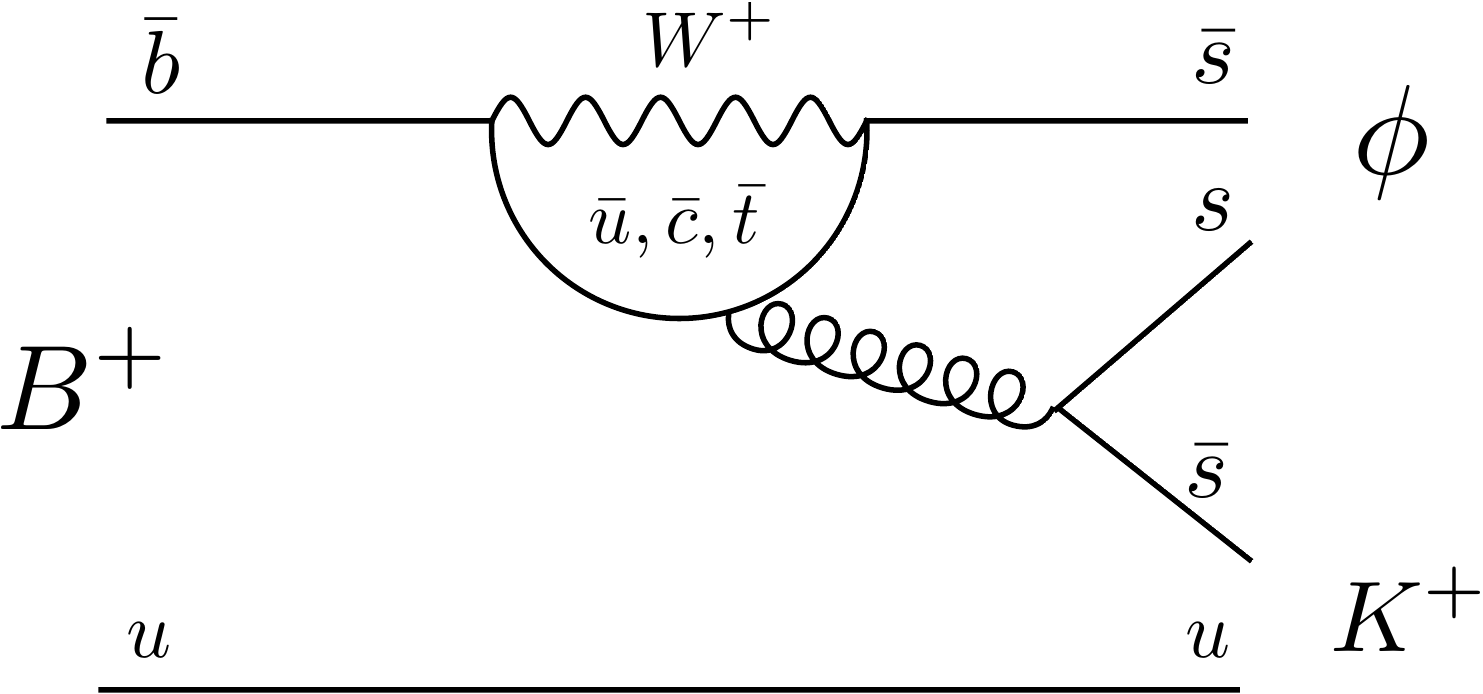}\hfill
\includegraphics[width=0.45\textwidth]{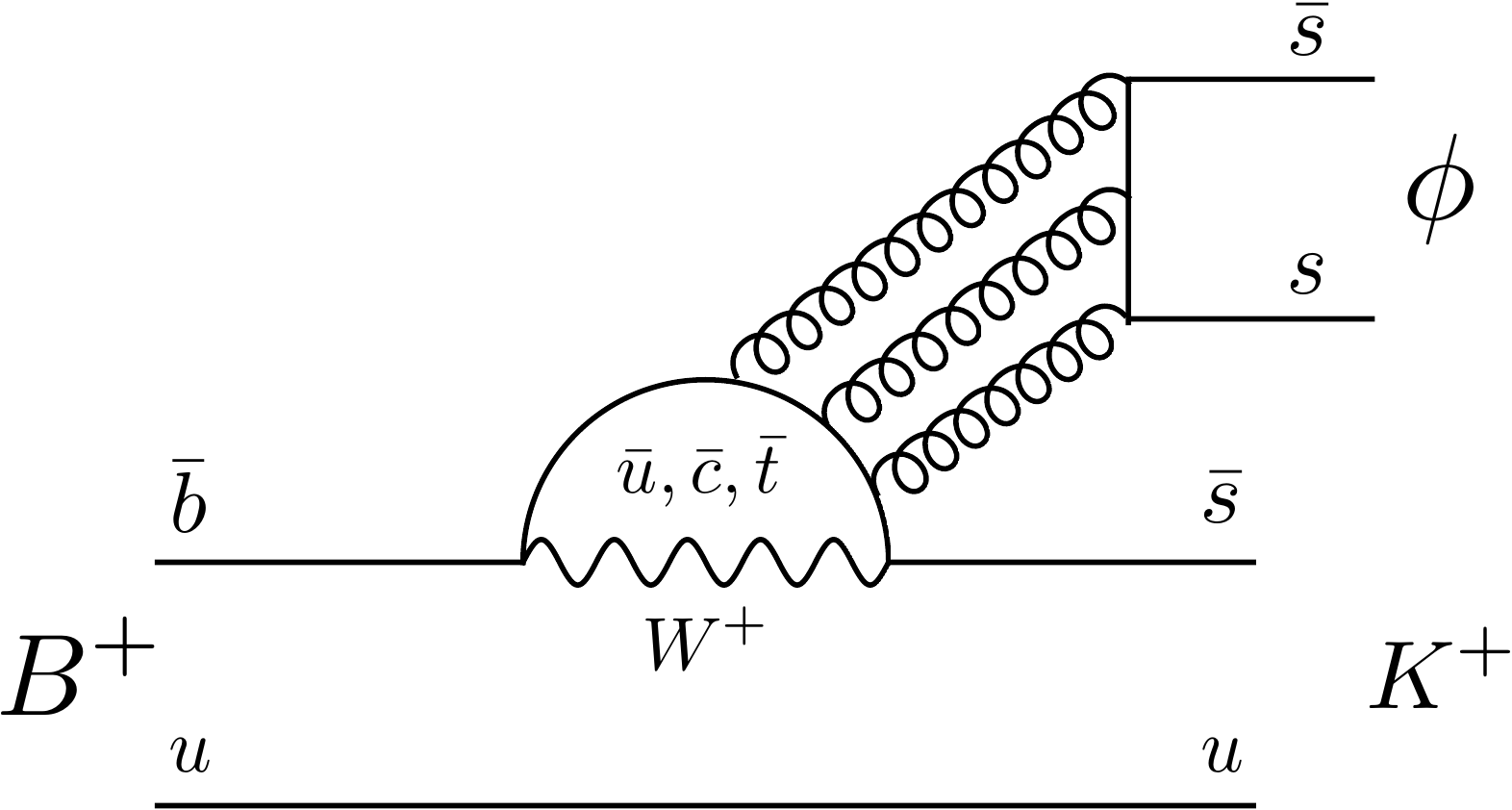}\\
\includegraphics[width=0.45\textwidth]{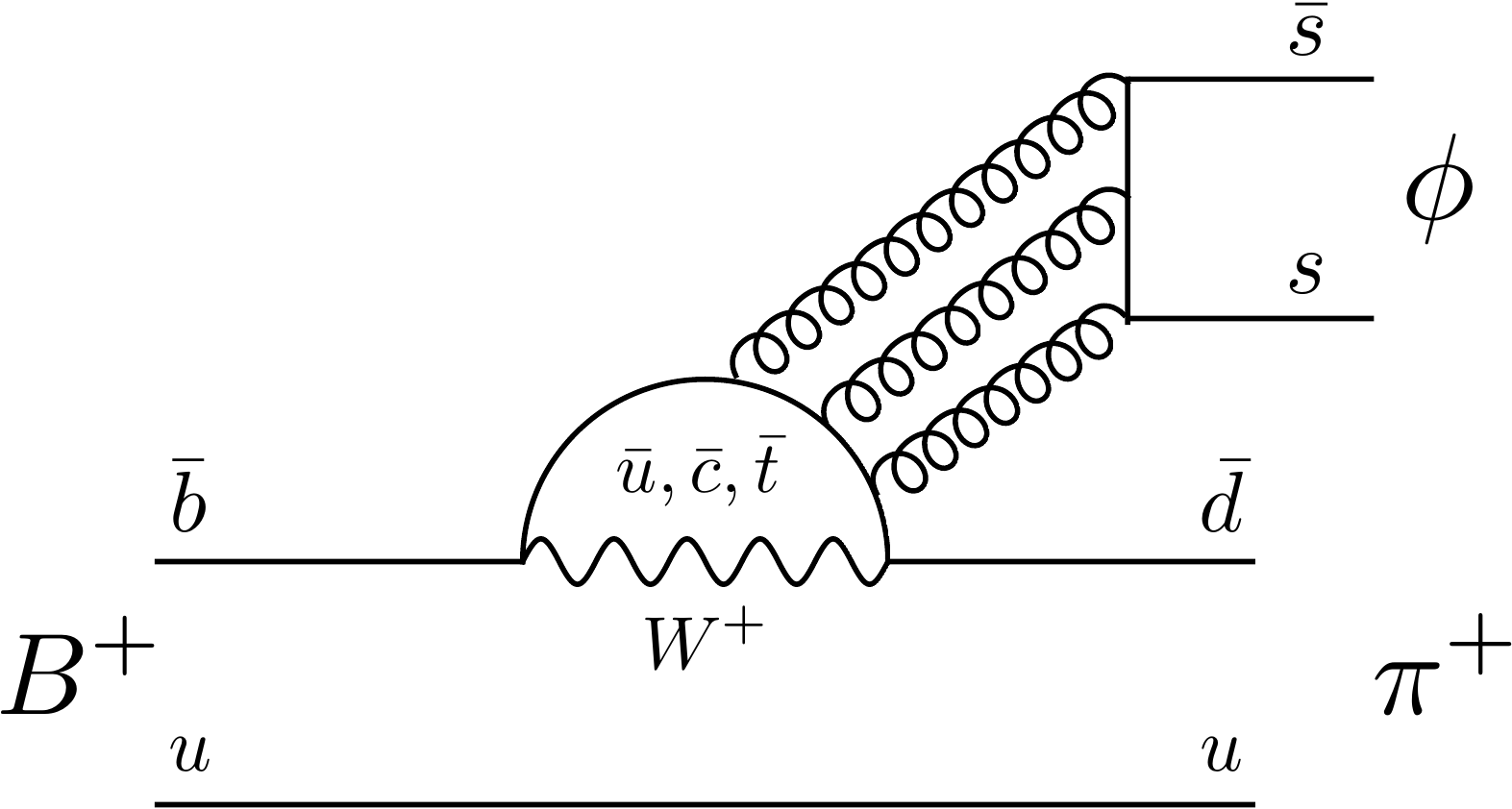}
\caption{\small Lowest-order Feynman diagrams of the Standard Model 
for the decays \decay{\Bp}{\phi\Kp} (top) and
\decay{\Bp}{\phi\pip} (bottom).
The diagrams with an external $\phi$ meson are OZI suppressed.}
\label{fig:FeynmanDiagram}
\end{figure}

The weak-interaction \decay{\Bpm}{\phi\Kpm} decay is governed by the
\decay{\bquark}{\squark\ssbar} transition. In the Standard Model (SM), 
it can only occur through loop diagrams
(see Fig.~\ref{fig:FeynmanDiagram}),
leading to a branching fraction of order $10^{-5}$~\cite{Beringer:1900zz}. 
Because the dominant amplitudes have similar weak phases, 
the 
\CP-violating charge asymmetry, defined as 
\begin{equation}
\ACP(\decay{\Bpm}{\phi \Kpm})\equiv 
\frac{\BR(\decay{\Bm}{\phi \Km})-\BR(\decay{\Bp}{\phi \Kp})}{\BR(\decay{\Bm}{\phi \Km})+\BR(\decay{\Bp}{\phi \Kp})} \,,
\end{equation}
is predicted to be small in the SM, typically 1--2\% with uncertainties of a few percent~\cite{Li:2006jv,Beneke:2003zv}.
A significantly larger value would signal interference with an amplitude 
not described in the SM. 
The current experimental world average is 
$\ACP(\mbox{\decay{\Bpm}{\phi \Kpm}})= 0.10 \pm 0.04$~\cite{Beringer:1900zz},
dominated by a recent measurement from the \babar collaboration~\cite{Lees:2012kxa}.
Large $CP$ violation effects have been seen in some regions of the \decay{\Bpm}{\Kp\Km\Kpm} phase space, but not around the $\phi$ resonance~\cite{LHCb-PAPER-2013-027}.

The \decay{\Bpm}{\phi\pipm} decay is another flavour-changing neutral current
process,
driven by the \decay{\bquark}{\dquark\ssbar} quark-level transition
(see Fig.~\ref{fig:FeynmanDiagram}).
The high suppression, due to the tiny product of the Cabibbo-Kobayashi-Maskawa
matrix elements~\cite{Cabibbo:1963yz,Kobayashi:1973fv} and to the Okubo-Zweig-Iizuka (OZI)
rule~\cite{Okubo:1963fa,Zweig:bothparts,Iizuka:1966fk}
associated with the creation of the colourless
$\ssbar$ pair forming the $\phi$ meson,
makes this rare loop decay a sensitive probe of the SM.
Indeed, even a small non-SM amplitude, \eg from R-parity violating supersymmetry~\cite{BarShalom:2002sv}, may dominate over the SM contribution. 

The current SM prediction for the \decay{\Bpm}{\phi\pipm} branching fraction
suffers from
uncertainties originating from the na\"ive factorization approach,
radiative corrections, calculation of the
long-distance contribution (\eg \decay{B}{K\Kstar} rescattering),
and $\omega-\phi$ mixing~\cite{Li:2009zj}. The latter is the main source of uncertainty. 
The physical $\omega$ and $\phi$ meson states do not coincide exactly with the ideal $(\ket{\uubar}+\ket{\ddbar})/\sqrt{2}$ and $\ket{\ssbar}$ states, respectively. They appear to be mixtures of these two states characterized by a small mixing angle $\delta_V$~\cite{Kucukarslan:2006wk,Benayoun:1999fv}, which depends on the magnitude of $\grpsuthree$ symmetry breaking and can be determined in the framework of chiral perturbation theory. However, more sophisticated treatments based on the full $\rho^0-\omega-\phi$ mixing scheme suggest that $\delta_V$ is mass dependent, i.e. takes different values at the $\omega$ and $\phi$ masses~\cite{Benayoun:2007cu,Benayoun:2009im}. In the QCD factorization approach, the  \decay{\Bpm}{\phi\pipm} branching fraction is predicted to be in the range $(5-10)\times10^{-9}$~\cite{Beneke:2003zv}  if $\omega-\phi$ mixing is neglected, but can be enhanced up to $0.6\times10^{-7}$~\cite{Gronau:2008kk,Li:2009zj} depending on the value of $\delta_V$.
However, the effect of $\omega-\phi$ mixing has not been observed in a recent search for $\decay{\Bd}{\jpsi\phi}$~\cite{Aaij:2013mtm}.
 Values of the \decay{\Bpm}{\phi\pipm} branching fraction in excess of $10^{-7}$ would be indicative of non-SM physics. 

The \decay{\Bpm}{\phi\pipm} decay mode has not been observed yet.
Currently, the most stringent experimental 
limit is 
$\BR(\decay{\Bpm}{\phi \pipm}) < 2.4\times10^{-7}$ at 90\% confidence level (CL), obtained
by the \babar collaboration~\cite{Aubert:2006nn}.

This Letter presents a measurement of the \decay{\Bpm}{\phi \Kpm} charge asymmetry and a search for the \decay{\Bpm}{\phi\pipm} decay mode
with the \lhcb detector.
The results are
based on a data sample collected during the 2011 $pp$ run of the Large Hadron Collider
at a centre-of-mass energy of 7\tev, corresponding to an integrated luminosity of 1.0\invfb. 
The $\phi$ meson is reconstructed in the $K^+K^-$ final state.
We define the $\phi$ signal as any peaking component in the $K^+K^-$ mass 
spectrum consistent with the known
parameters of the $\phi$ resonance, without attempting a full amplitude
analysis of the three-body $K^+K^-K^\pm$ and $K^+K^-\pi^\pm$ final
states.
In order to suppress several systematic
effects, the primary observables measured in this analysis are the difference of $\CP$-violating charge asymmetries
\begin{equation}
\Delta \ACP \equiv  \ACP(\decay{\Bpm}{\phi \Kpm})-\ACP(\decay{\Bpm}{\jpsi\Kpm}) \,, 
\label{eq:Delta_ACP}
\end{equation}
and the branching fraction ratio
$\BR(\decay{\Bpm}{\phi\pipm})/\BR(\decay{\Bpm}{\phi\Kpm})$, which are then converted to results on  $\ACP(\decay{\Bpm}{\phi \Kpm})$ and $\BR(\decay{\Bpm}{\phi\pipm})$
using the best known values of $\ACP(\decay{\Bpm}{\jpsi\Kpm})$~\cite{Beringer:1900zz,Abazov:2013sqa} and $\BR(\decay{\Bpm}{\phi\Kpm})$~\cite{Beringer:1900zz}.
The choice of \decay{\Bpm}{\jpsi\Kpm} as reference channel and other features of the analysis follow the approach adopted in inclusive studies of \decay{\Bpm}{\Kp\Km\Kpm} decays
with the same data set~\cite{LHCb-PAPER-2013-027}. 

The two measurements are performed in a common analysis, \ie they are based 
on identical event selections and data descriptions whenever possible. 
The observables are obtained from two-dimensional maximum likelihood fits to
the unbinned \Bpm and $\phi$ mass distributions of the reconstructed candidates,
using parametric shapes with minimal dependence on simulation.
The results of these fits were not examined until
the entire analysis procedure was finalized. 

\section{Detector and data set}
\label{sec:Detector}

The \lhcb detector~\cite{Alves:2008zz} is a single-arm forward
spectrometer covering the \mbox{pseudorapidity} range $2<\eta <5$,
designed for the study of particles containing \bquark or \cquark
quarks. The detector includes a high-precision tracking system
consisting of a silicon-strip vertex detector surrounding the $pp$
interaction region, a large-area silicon-strip detector located
upstream of a dipole magnet with a bending power of about
$4{\rm\,Tm}$, and three stations of silicon-strip detectors and straw
drift tubes placed downstream.
The combined tracking system provides a momentum measurement with a
relative uncertainty that varies from 0.4\% at 5\gevc to 0.6\% at 100\gevc,
and an impact parameter (IP) resolution of 20\mum for
tracks with high transverse momentum (\pt). Charged hadrons are identified
using two ring-imaging Cherenkov detectors~\cite{LHCb-DP-2012-003}.
Photon, electron and
hadron candidates are identified by a calorimeter system consisting of
scintillating-pad and preshower detectors, an electromagnetic
calorimeter and a hadronic calorimeter. Muons are identified by a
system composed of alternating layers of iron and multiwire
proportional chambers. 
The direction of the magnetic field of the spectrometer dipole magnet is reversed regularly. 

The trigger~\cite{LHCb-DP-2012-004} consists of a
hardware stage, based on information from the calorimeter and muon
systems, followed by a software stage, which applies a full event
reconstruction. The \Bpm candidate decays considered in this analysis must belong to one of two exclusive categories of events,
called TOS (triggered on signal) or TIS (triggered independently of signal).
A TOS event is triggered at the hardware stage by one of the candidate's final-state particles being compatible with a transverse energy deposit greater than 3.5\gev in the hadron calorimeter. 
A TIS event does not satisfy the TOS definition and is triggered
at the hardware stage by activity in the rest of the event. 
All candidates must pass a software trigger requiring
a \mbox{two-,} three- or four-track
secondary vertex with a large scalar
sum of the transverse momentum of
the tracks and a significant displacement from the primary $pp$
interaction vertices~(PVs). 
 At least one track should have $\pt >1.7\gevc$ and \chisqip with respect to any
PV greater than 16, where \chisqip is defined as the
difference in \chisq of a given PV reconstructed with and
without the considered track. A multivariate algorithm~\cite{BBDT} is used for
the identification of secondary vertices consistent with the decay
of a \bquark hadron.

In the simulation, $pp$ collisions are generated using
\pythia~6.4~\cite{Sjostrand:2006za} with a specific \lhcb
configuration~\cite{LHCb-PROC-2010-056}.  Decays of hadronic particles
are described by \evtgen~\cite{Lange:2001uf}, in which final state
radiation is generated using \photos~\cite{Golonka:2005pn}. The
interaction of the generated particles with the detector and its
response are implemented using the \geant
toolkit~\cite{Allison:2006ve, *Agostinelli:2002hh} as described in
Ref.~\cite{LHCb-PROC-2011-006}.

\section{Event selection and efficiency}
\label{sec:Event_selection}

The selections of \decay{\Bpm}{\phi\Kpm} and \decay{\Bpm}{\phi\pipm}
candidates
are identical, except for the particle identification (PID) requirement
on the charged hadron combined with the $\phi$ candidate, 
which is referred to as the bachelor hadron
$h^{\pm}$ ($h^{\pm} = \Kpm$ or \pipm). 
The other requirements are chosen to minimize the relative statistical uncertainty on the \decay{\Bpm}{\phi\Kpm}  signal yield.

Only good quality tracks with $\chisqip > 25$ and $\pt > 0.25\gevc$ are used
in the reconstruction.
The $\phi$ meson candidates are reconstructed from two oppositely-charged tracks
identified as kaons with the PID requirement ${\rm DLL}_{K\pi} >2$, 
where ${\rm DLL}_{K\pi}$ is the difference in log-likelihood between
the kaon and pion hypotheses, as determined with the ring-imaging Cherenkov detectors in control samples of known particle composition~\cite{LHCb-DP-2012-003}. The $\phi$ candidates are
required to have $\pt > 2\gevc$, a total momentum, $p$, larger than 10\gevc
and an invariant mass, $m_{KK}$, in the range 1.00--1.05\gevcc.
Bachelor hadrons, reconstructed either as pions if ${\rm DLL}_{K\pi} <-1$
or kaons otherwise, are required to have $p> 10\gevc$ and $\pt > 2.5\gevc$, 
and are combined with $\phi$ candidates to form
\decay{\Bpm}{\phi h^{\pm}} candidates. 
These \Bpm candidates are required to have $\pt > 2\gevc$,
a three-track vertex $\chi^2$ per degree of freedom less than 9, and an 
invariant mass $m_{KKh}$ in the range 5.0--5.5\gevcc. Furthermore
$\cos\theta_{\rm p}$ is required to be greater than 0.9999,
where $\theta_{\rm p}$ is the angle between the \Bpm momentum vector
and the vector joining the \Bpm production vertex to the \Bpm decay vertex. The production vertex is chosen as the PV 
for which the \Bpm has the smallest $\chisqip$.

Multiple candidates, occurring in 0.2\% of the events, are removed by keeping the candidate with the smallest $\Bpm$ vertex $\chi^2$. The final data sample consists of 6251  \decay{\Bpm}{\phi\Kpm}  candidates and 2169 \decay{\Bpm}{\phi\pipm} candidates.

The PID performance is determined from a large and high-purity sample of
pions from prompt $\decay{\Dstarp}{\Dz(\Km\pip)\pip}$ and $\decay{\Dstarm}{\Dzb(\Kp\pim)\pim}$ decays,
as a function of $p$ and $\eta$. 
After reweighting this calibration sample to the same momentum and 
pseudorapidity distributions as for the  bachelor pion in simulated 
\decay{\Bpm}{\phi\pipm} decays,
the efficiency of the
PID requirement ${\rm DLL}_{K\pi} <-1$ for the bachelor pion 
is measured to be $\rm 0.846 \pm 0.011\, (stat) \pm 0.020\, (syst)$, with a 5\% kaon misidentification probability.
All other efficiencies, which are slightly different for
\decay{\Bpm}{\phi\pipm} and \decay{\Bpm}{\phi\Kpm} decays due to their kinematic properties, 
are determined from simulation. The efficiency ratio 
\begin{equation}
\frac{\epsilon(\decay{\Bpm}{\phi\pipm})}{\epsilon(\decay{\Bpm}{\phi\Kpm})} =
\rm 0.762 \pm 0.031 \,(stat) \pm 0.036 \,(syst)
\label{eq:eff_ratio}
\end{equation}
is obtained, where the numerator is the total efficiency for a
\decay{\Bpm}{\phi\pipm} decay to be selected as a \decay{\Bpm}{\phi\pipm}
candidate and the denominator is the total efficiency for a
\decay{\Bpm}{\phi\Kpm} decay to be selected either as a
\decay{\Bpm}{\phi\Kpm} candidate or as a \decay{\Bpm}{\phi\pipm} candidate.
The statistical uncertainty arises from the size of the calibration and simulation
samples, while the systematic uncertainty is the quadratic sum of
contributions from the PID ($\pm 0.018$), the trigger ($\pm 0.008$),
and other offline kinematic selection requirements ($\pm 0.030)$.

\section{Fit description}
\label{sec:Fit_description}

The observables of interest, namely the 
asymmetry between the \decay{\Bm}{\phi\Km}
and \decay{\Bp}{\phi\Kp} yields and 
the ratio between the \decay{\Bpm}{\phi\pipm} and 
\decay{\Bpm}{\phi\Kpm} yields, are each determined
from a two-dimensional unbinned extended maximum likelihood fit based on 
probability density functions (PDFs) of the $m_{KKh}$ and $m_{KK}$ masses.
In each case, 
independent subsamples of events, each with either \decay{\Bpm}{\phi\Kpm} 
candidates or \decay{\Bpm}{\phi\pipm}
candidates, 
are fitted simultaneously. For each subsample, 
the likelihood is written as 
\begin{equation}
{\cal L} = {\textstyle \exp(-\sum_j N_j)} \, \prod_i^N
{\left( \textstyle \sum_j N_j P_j^i\right)} \,,
\end{equation}
where $N_j$ is the yield of fit component~$j$, $P_j^i$ is the probability
of event $i$ for component~$j$, and the index $i$ runs over the $N$ events
in the subsample. Except for the misidentified components described further below, the probabilities $P_j^i$ are given by the product of
PDFs for the two $K^+K^-h^{\pm}$ and $K^+K^-$ invariant masses,
evaluated at the values $m_{KKh}^i$ and $m_{KK}^i$ of event $i$:
\begin{equation}
P_j^i = P_j^{KKh}(m_{KKh}^i) P_j^{KK}(m_{KK}^i) \,.
\end{equation}
This assumes that the two mass variables are independent,
as supported by data and simulation studies. The
correlation between $m_{KKh}$ and $m_{KK}$ is found to be less than $4\%$.

The description of the $m_{KKh}$ distributions involves a combination of
three contributions: a signal peaking at the \Bpm mass,
a broad low-mass background with an end-point near $5150\mevcc$ due to
partially-reconstructed \bquark-hadron decays such as \decay{\Bd}{\phi\Kstarz},
and a linear background from random combinations.
The peaking signal is modelled with a Crystal Ball
function~\cite{Skwarnicki:1986xj} modified such that both the upper and lower tails are power laws.
The mean and the width $\sigma_B$ of the Crystal Ball function are free in the fit,
while the tail parameters are determined from simulation.
The partially-reconstructed background is described with an ARGUS function~\cite{Argus}
convoluted with a Gaussian resolution function
of the same $\sigma_B$ as the \Bpm signal.
The $m_{KK}$ distribution is described with two contributions:
a peaking term centred on the $\phi$ mass, described with
a relativistic Breit-Wigner function convoluted with a Gaussian
resolution function of free width,
and a linear term originating from nonresonant, S-wave,
or random combinations of two kaons.
The above three $m_{KKh}$ contributions and two $m_{KK}$ contributions lead 
to six components for each subsample:
the \decay{\Bpm}{\phi h^\pm} signal, the nonresonant 
\decay{\Bpm}{\Kp\Km h^\pm} background, the partially-reconstructed 
$b$-hadron backgrounds with or without a true $\phi$ meson 
(for example \decay{B}{\phi h^\pm \pi} or \decay{B}{\Kp\Km h^\pm \pi}), 
and the combinatorial backgrounds with or without a true  $\phi$ meson. 
The nonresonant \decay{B^{\pm}}{\Kp\Km h^\pm} components include $b\rightarrow c$ decays, which are found to be negligible from simulation studies.
%Reflections from $b\rightarrow c$ decays included in the nonresonant \decay{B}{\Kp\Km h^\pm} components are found to be negligible from studies on simulation data.

In addition, we consider
components for the misidentified \decay{\Bpm}{\phi\Kpm} and
\decay{\Bpm}{\Kp\Km\Kpm} decays in the \decay{\Bpm}{\phi\pipm} sample,
while misidentified \decay{\Bpm}{\phi\pipm} and \decay{\Bpm}{\Kp\Km\pipm}
decays in the \decay{\Bpm}{\phi\Kpm} sample are negligible,
and therefore ignored.
For these two additional components, the $m_{KK\pi}$ PDF is conditional to the observable $\delta m = m_{KKK} - m_{KK\pi}$, which is the mass difference under the two bachelor hadron mass hypotheses. The probabilities are written as
\begin{equation}
P_j^i = P_{\rm misID}^{KK\pi}(m_{KK\pi}^i|\delta m^i) P_j^{KK}(m_{KK}^i) \,,
\end{equation}
where $P_j^{KK}$ is the $m_{KK}$ PDF described above (representing 
either $\phi$ signal or background) and 
\begin{equation}
\left.
P_{\rm misID}^{KK\pi} (m^i_{KK\pi}| \delta m^i)=
P_{\phi K}^{KKK} (m^i_{KK\pi}+\delta m^i)
\right|_{\sigma_B \to \rho\sigma_B} \,.
\label{eq:misid}
\end{equation}
Here $P_{\phi K}^{KKK}$ is the $m_{KKK}$ PDF of the \decay{\Bpm}{\phi\Kpm} signal,
but with an increased \Bpm mass resolution
to account for the effects of the typically
higher momentum of misidentified bachelor kaons.
The parameter $\sigma_B$ is multiplied here by the central value of a 
factor $\rho=1.26\pm 0.10$, determined from data as the ratio of
the measured $m_{KKK}$ resolutions of the \decay{\Bpm}{\phi\Kpm} signal
in the regions $-7 < {\rm DLL}_{K\pi} <-1$ and ${\rm DLL}_{K\pi} > -1$.
The expression in Eq.~\ref{eq:misid} is equivalent to 
$\left. P_{\phi K}^{KKK} (m^i_{KKK})
\right|_{\sigma_B \to \rho\sigma_B}$, 
which means that the \decay{\Bpm}{\phi\Kpm} misidentified component 
in the \decay{\Bpm}{\phi\pipm} sample would have a 
\decay{\Bpm}{\phi\Kpm} signal distribution if the correct mass was assigned to the 
bachelor kaon. 
The advantage of introducing the $\delta m$ observable is to connect
the \decay{\Bpm}{\phi\Kpm} shapes in the \decay{\Bpm}{\phi\pipm} and 
\decay{\Bpm}{\phi\Kpm} samples, thereby constraining the misidentified 
\decay{\Bpm}{\phi\Kpm} component in the \decay{\Bpm}{\phi\pipm} sample
using the large signal in the \decay{\Bpm}{\phi\Kpm} sample. This procedure allows to describe the misidentified component with the same parametric shape as the \decay{\Bpm}{\phi\pipm} signal, and reduces the 
statistical uncertainty on the \decay{\Bpm}{\phi\pipm} yield by a factor of two. 
However, this introduces a bias
because the $\delta m$ distribution, which is not 
accounted for in the likelihood, is not the
same for all components~\cite{Punzi:2004wh}.
To reduce this bias, the \decay{\Bpm}{\phi\pipm} sample is divided
into four bins of $\delta m$, 
each with its own eight components. 
This procedure reduces the bias on the
\decay{\Bpm}{\phi\pipm} signal yield to a negligible level. 

Other fit parameters that are common to the different subsamples 
are the $m_{KKh}$ end-point of the partially-reconstructed backgrounds, the peaking $m_{KK}$ PDF parameters for all components containing a $\phi$ meson, and the $m_{KK}$ slope of the nonresonant \decay{\Bpm}{\Kp\Km h^{\pm}} components. Finally, the ratio of the yield of the misidentified nonresonant \decay{\Bpm}{\Kp\Km\Kpm} background to the yield of misidentified \decay{\Bpm}{\phi\Kpm} background in the \decay{\Bpm}{\phi\pipm} sample is constrained to the yield ratio of the corresponding correctly-identified components in the \decay{\Bpm}{\phi\Kpm} sample.

The fit procedure is validated on
simulated data containing the expected proportion of signal and background
events. 

These studies, which take into account the 
different $\delta m$ distributions and the possible correlation 
between the fit observables, demonstrate the stability of the fit
and show that the fit results are unbiased.

\section{Measurement of the \decay{\Bpm}{\phi\Kpm} charge asymmetry}
\label{sec:ACP}

The charge asymmetry of the \decay{\Bpm}{\phi\Kpm} signal is determined
from a fit to the 
\decay{\Bm}{\phi\Km} and \decay{\Bp}{\phi\Kp} candidates in the 
${\rm DLL}_{K\pi} > -1$ region. These two samples are each divided into 
two subsamples according to whether the events were TOS or TIS
at the hardware trigger stage.
In the fit, each of the six components has therefore four yields.
For the signal component, they are expressed as 
$N^{\pm}_{\rm TOS} = N_{\rm TOS} (1\mp \A_{\rm raw,TOS})/2$ and
$N^{\pm}_{\rm TIS} = N_{\rm TIS} (1\mp \A_{\rm raw,TIS})/2$, 
where $N_k$ is the total yield and $\A_{{\rm raw},k}$ is the 
raw yield asymmetry in subsample $k$ ($k = \rm TOS, TIS$).
The fit has a total of 34 free parameters: 10 mass shape parameters, 
12 yields and 12 raw asymmetries.

\begin{figure}[t]
\centering
\begin{overpic}[width=0.48\textwidth]{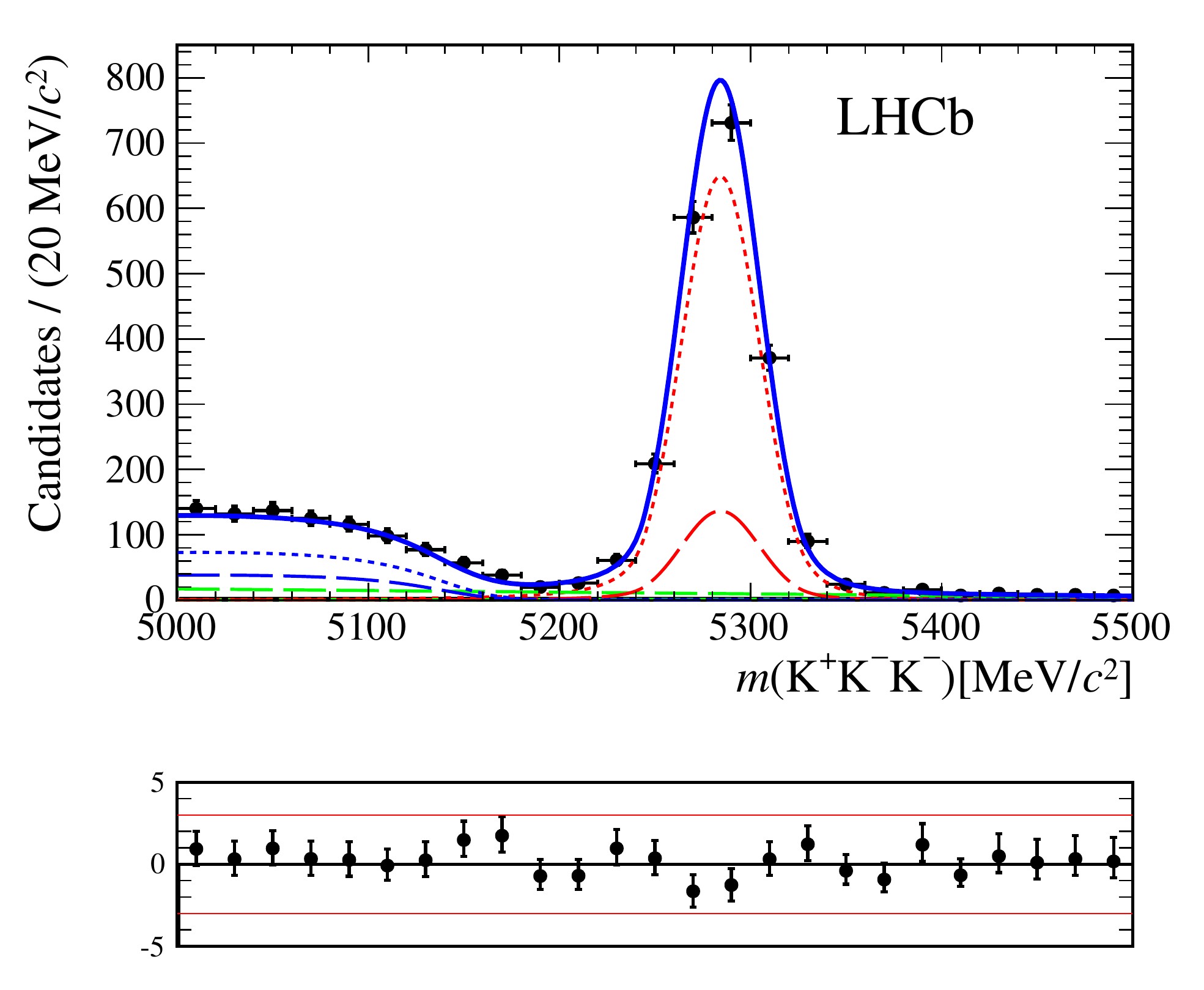}%
\put(85,73){\scriptsize (a)}
\end{overpic}
\begin{overpic}[width=0.48\textwidth]{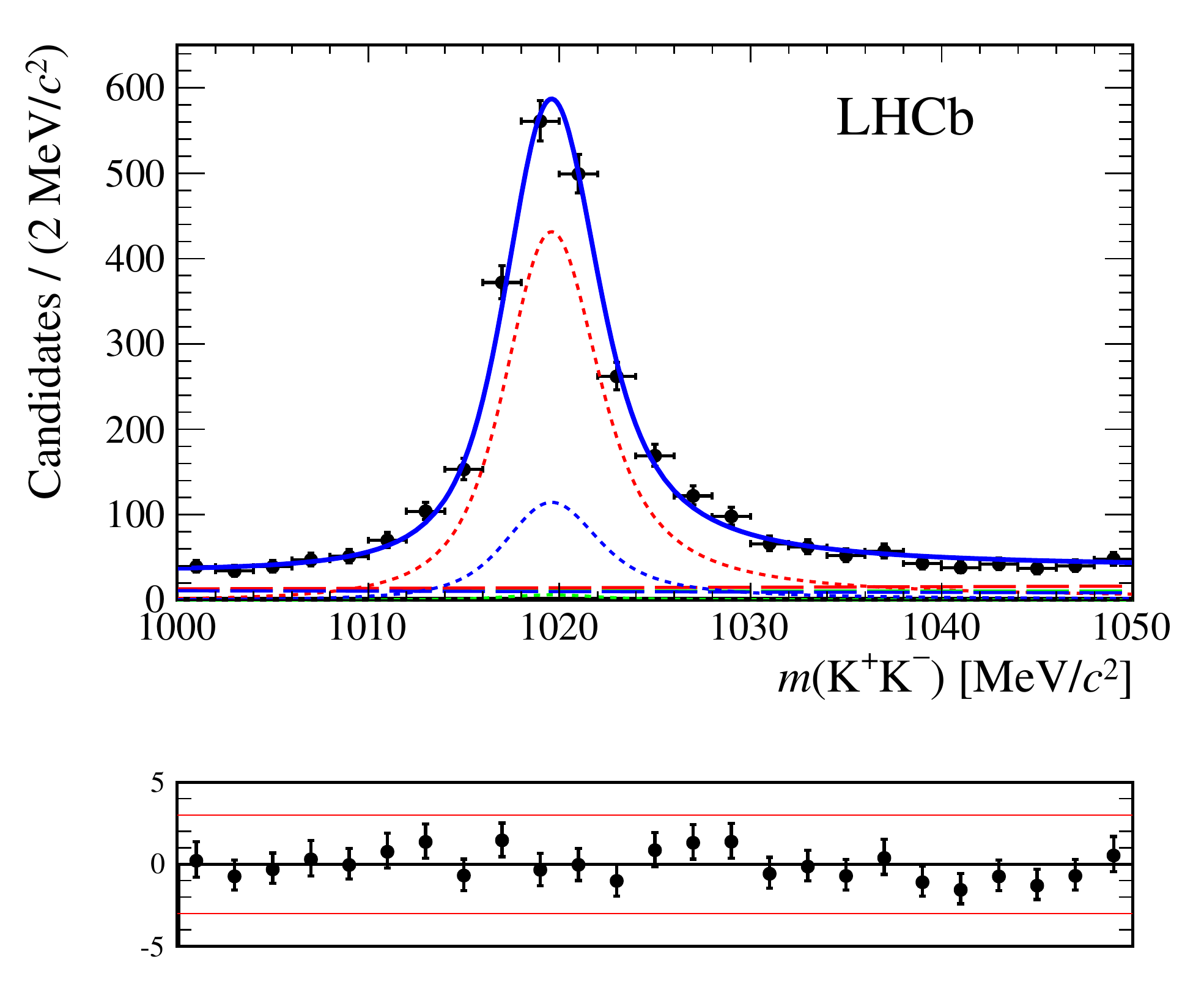}
\put(85,73){\scriptsize (b)}
\end{overpic}
\begin{overpic}[width=0.48\textwidth]{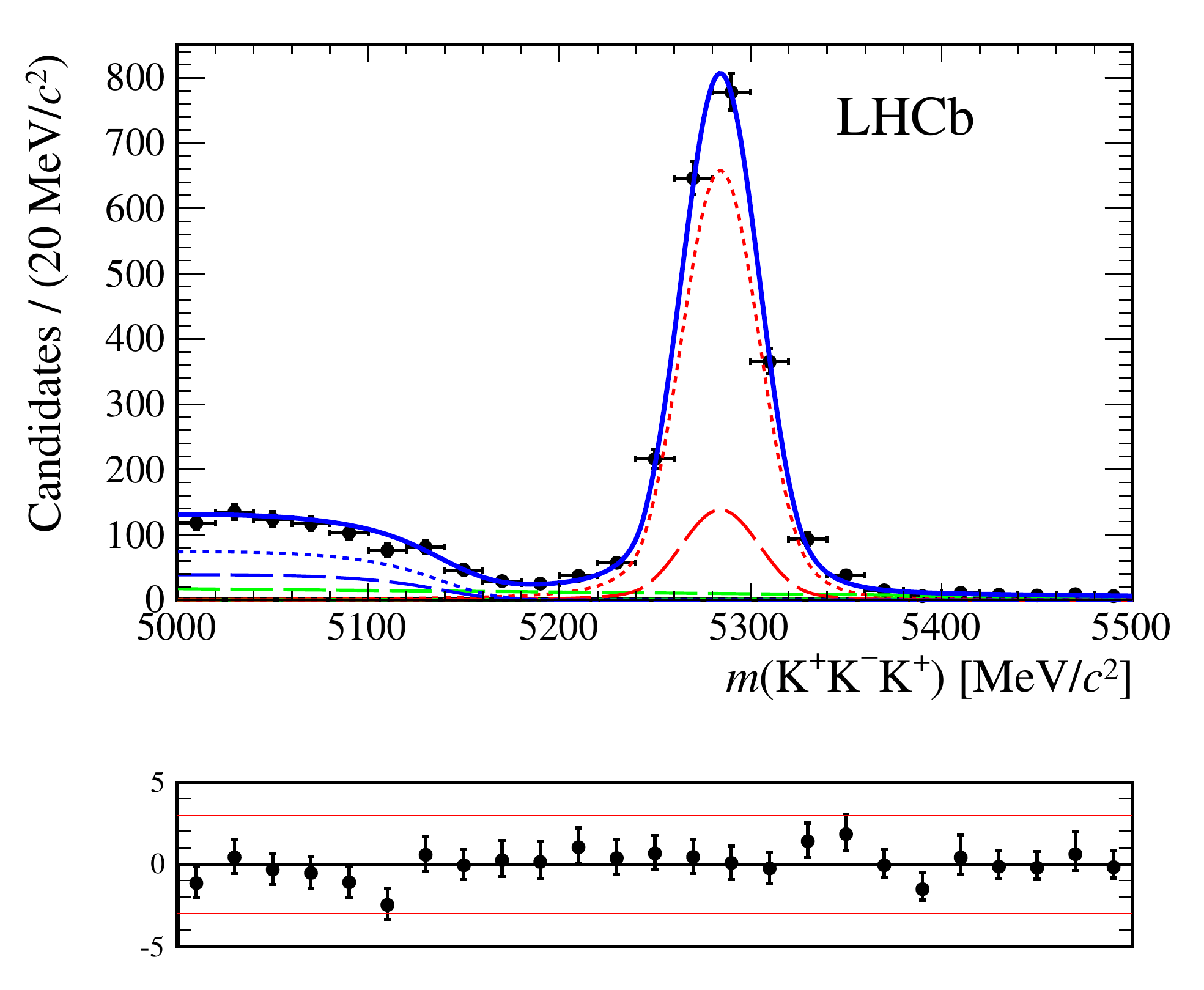}%
\put(85,73){\scriptsize (c)}
\end{overpic}
\begin{overpic}[width=0.48\textwidth]{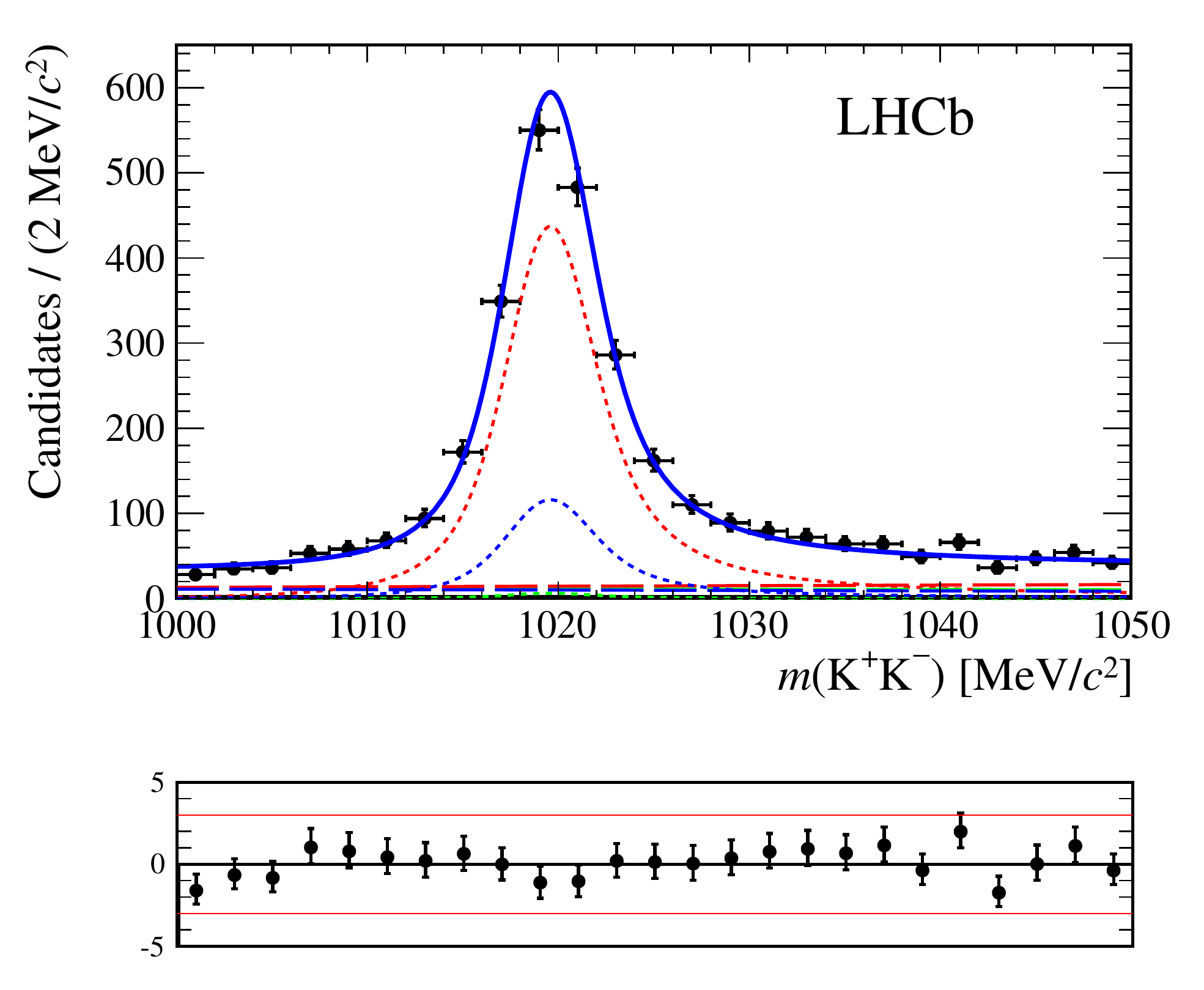}
\put(85,73){\scriptsize (d)}
\end{overpic}
\caption{\small
Distributions of the (a) $\Kp\Km\Km$ and (b) $\Kp\Km$ masses of
the selected \decay{\Bm}{\phi\Km} candidates, as well as of the
(c) $\Kp\Km\Kp$ and (d) $\Kp\Km$ masses of
the selected \decay{\Bp}{\phi\Kp} candidates.
The solid blue curves represent the result
of the simultaneous fit described in the text, with the following components:
\decay{\Bpm}{\phi\Kpm} signal (dotted red), 
nonresonant \decay{\Bpm}{\Kp\Km\Kpm} background (dashed red),
partially-reconstructed $b$-hadron background with (dotted blue) or without 
(dashed blue) a true $\phi$ meson,
and combinatorial background with (dotted green) or without (dashed green) a true $\phi$ meson.
Some of the components are barely visible because the corresponding yields are small.
Normalized residuals are displayed below each histogram.}
\label{fig:PhiKAsymmetry}
\end{figure}

Figure~\ref{fig:PhiKAsymmetry} shows the projections
of the fitting function superimposed on the  $m_{KKK}$ and $m_{KK}$
distributions, shown 
separately for $B^-$ and $B^+$ candidates, but where TOS and TIS events are summed. 
The $m_{KKK}$ resolution measured from the fit is $\sigma_B=20.4\pm0.3\mevcc$.
The fitted raw asymmetries for the signal
are shown in the first line of Table~\ref{tab:Asymmetries}. They are statistically uncorrelated. 

\begin{table}
\caption{\small Raw charge asymmetries for the
\decay{\Bpm}{\phi \Kpm} and \decay{\Bpm}{\jpsi \Kpm} decays, 
their difference $\Delta \ACP$, and the fraction of 
\decay{\Bpm}{\phi \Kpm} signal events in each trigger subsample for $k = \rm TOS, TIS$.
All uncertainties are statistical only.}
\begin{center}\begin{tabular}{lcc}
  & TOS subsample & TIS subsample \\
\hline  
$\A_{{\rm raw},k}(\decay{\Bpm}{\phi\Kpm})$
  & $+0.027\pm 0.026$ & $-0.053\pm 0.035$ \\
$\A_{{\rm raw},k}(\decay{\Bpm}{\jpsi\Kpm})$ 
  & $-0.024\pm 0.008$ & $-0.008\pm 0.005$ \\
$\Delta\ACP$
  & $+0.052\pm 0.027$ & $-0.045\pm 0.035$ \\
$N_k/(N_{\rm TOS}+N_{\rm TIS})$ 
  & 66\% & 34\% \\
\hline
Weighted $\Delta\ACP$ average 
  & \multicolumn{2}{c}{$+0.019\pm 0.021$} \\
\end{tabular}\end{center}
\label{tab:Asymmetries}
\end{table}

Each raw charge asymmetry is related to the \CP asymmetry through
\begin{equation}
\A_{{\rm raw},k}(\decay{\Bpm}{\phi \Kpm}) = 
\ACP(\decay{\Bpm}{\phi \Kpm}) +
{\cal A}_{{\rm D},k}(\decay{\Bpm}{\phi \Kpm}) + {\cal A}_{\rm P} \,,
\label{eq:raw}
\end{equation}
where ${\cal A}_{{\rm D},k}(\decay{\Bpm}{\phi \Kpm})$ is the
detection charge asymmetry for the bachelor \Kpm 
and ${\cal A}_{\rm P}$ is the production asymmetry of \Bpm mesons.
Equation~\ref{eq:raw} and the corresponding equation for the \decay{\Bpm}{\jpsi\Kpm} reference channel hold because all involved asymmetries are small. 
Under the assumption that the detection asymmetry
is the same for \decay{\Bpm}{\phi \Kpm} and \decay{\Bpm}{\jpsi\Kpm}, 
which is correct in the limit where the bachelor \Kpm has
the same kinematic properties,
the difference in charge asymmetries
defined in Eq.~\ref{eq:Delta_ACP} can be written as 
\begin{equation}
\Delta \ACP = \A_{{\rm raw},k}(\decay{\Bpm}{\phi \Kpm})-\A_{{\rm raw},k}(\decay{\Bpm}{\jpsi\Kpm}) 
\label{eq:assumption}
\end{equation}
and should not depend on the trigger category $k$.
The raw charge asymmetries of \decay{\Bpm}{\jpsi\Kpm} decays have been 
measured in a previous analysis~\cite{LHCb-PAPER-2013-027};
they are subtracted from the 
\decay{\Bpm}{\phi \Kpm} raw asymmetries
to obtain two independent measurements of $\Delta \ACP$.
Since the two results agree within about two statistical standard deviations, 
the results are combined. 
The final $\Delta \ACP$ result is computed as a weighted average,
with weights equal to the fractions $N_k/(N_{\rm TOS}+N_{\rm TIS})$ of
signal events in the two trigger subsamples.
All inputs to the calculation are reported in Table~\ref{tab:Asymmetries}. 
The separation between TIS and TOS events is needed because the 
detection asymmetry ${\cal A}_{{\rm D},k}$ depends on the trigger category $k$ and 
the fraction of events in the two categories differs between the  
signal and reference channels.

\begin{table}[t]
\caption{\small Systematic uncertainties on the measurement of $\Delta\ACP$.}
\begin{center}
\begin{tabular}{lc}
Source & Uncertainty \\
\hline
Mass shape modelling          & 0.003 \\ 
Possible S-wave contribution  & 0.002 \\ 
Trigger                       & 0.004 \\ 
Bachelor kaon kinematic properties & 0.005 \\ 
Geometric acceptance        & 0.002 \\ 
\hline
Quadratic sum                 & 0.007 \\ 
\end{tabular}\end{center}
\label{tab:SystematicsACP}
\end{table}

Several systematic uncertainties are considered
on the weighted $\Delta \ACP$ average,
as summarized in Table~\ref{tab:SystematicsACP}.
The contribution due to the mass shape modelling is 
obtained by repeating the fit
(and the calculation of Table~\ref{tab:Asymmetries})
with the fixed parameter values of the Crystal Ball and ARGUS functions changed within their uncertainties,
as determined from simulation and $\decay{\Bpm}{\phi \Kpm}$ data, respectively,
or with an exponential (rather than linear) combinatorial background model. 
%The possible effect of an S-wave component on the signal yield
%is estimated in Sec.~\ref{sec:BR}; if this would correspond
%to an additional component included in the signal without charge asymmetry,
%a bias would appear on $\Delta\ACP$, which is taken as a systematic uncertainty.
Possible residual effects from S-wave contributions not fully accounted for by the linear component are
investigated by comparing the observed angular distribution of the \decay{\Bpm}{\phi\Kpm} signal 
with the expectation for a peaking structure in the $K^+K^-$ mass due to a single 
P-wave state. Other P-wave components are neglected. If these S-wave contributions corresponded
to an additional component included in the signal without charge asymmetry,
a bias would appear on $\Delta\ACP$, which is taken as a systematic uncertainty.

The charge asymmetry in the trigger efficiency for kaons
of the TOS subsample does not completely cancel in $\Delta\ACP$, because 
of the different number of kaons in the two decay modes considered.
The difference between the values of
$\A_{\rm raw,TOS}(\decay{\Bpm}{\jpsi\Kpm})$ computed
with and without a charge-dependent correction for the kaon efficiency
determined from calibration data
is propagated as a systematic uncertainty on $\Delta\ACP$.
Such an effect is absent for the TIS subsample.
Another small contribution, due to the TOS events that would still be accepted by the hardware trigger level without considering
the particles from the $\Bp$ candidate decay, has been included in the trigger systematic uncertainty.
Due to differences in the kinematic selections of the \decay{\Bpm}{\phi\Kpm}
and \decay{\Bpm}{\jpsi\Kpm} decay modes,
the assumption of Eq.~\ref{eq:assumption} cannot be exact,
and a further systematic uncertainty is assigned. 
The fit of the raw charge asymmetries of \decay{\Bpm}{\jpsi\Kpm}
is repeated with the same kinematic selection on the bachelor kaon 
as for \decay{\Bpm}{\phi \Kpm},
\ie $p>10\gevc$ and $\pt > 2.5\gevc$, and after reweighting its momentum 
distribution to that observed in the \decay{\Bpm}{\phi \Kpm} decays.
The resulting effect on
$\Delta\ACP$ is taken as a systematic uncertainty. 
Finally, we repeat the \decay{\Bpm}{\phi \Kpm}
analysis after requiring the bachelor kaon momentum 
to point in a fiducial solid angle avoiding detector edge effects, 
and assign the observed change in $\Delta\ACP$ as a systematic uncertainty 
due to the geometrical acceptance. 

The final measurement is 
\begin{equation}
\Delta\ACP = \rm 0.019\pm 0.021 (stat) \pm 0.007 (syst) \,.
\end{equation}
A recent update of the \decay{\Bpm}{\jpsi\Kpm} charge asymmetry measurement by the 
D0 collaboration~\cite{Abazov:2013sqa}
has not been included yet in the average of 
the Particle Data Group (PDG)~\cite{Beringer:1900zz}. 
Replacing the previous D0 result with the new one yields
the world average $\ACP(\decay{\Bpm}{\jpsi\Kpm}) = 0.003 \pm 0.006$,
where the uncertainty is scaled by a factor 1.8
according to the PDG averaging rules. 
Using this average, we obtain 
\begin{equation}
\ACP(\decay{\Bpm}{\phi \Kpm}) = \rm 0.022\pm 0.021 (stat) \pm 0.009 (syst) \,,
\end{equation}
where the uncertainty on the \decay{\Bpm}{\jpsi\Kpm} charge asymmetry is 
incorporated in the systematic uncertainty.

\section{Search for \decay{\Bpm}{\phi\pipm} decays}
\label{sec:BR}
\begin{figure}[t!]
\centering
\begin{overpic}[width=0.48\textwidth]{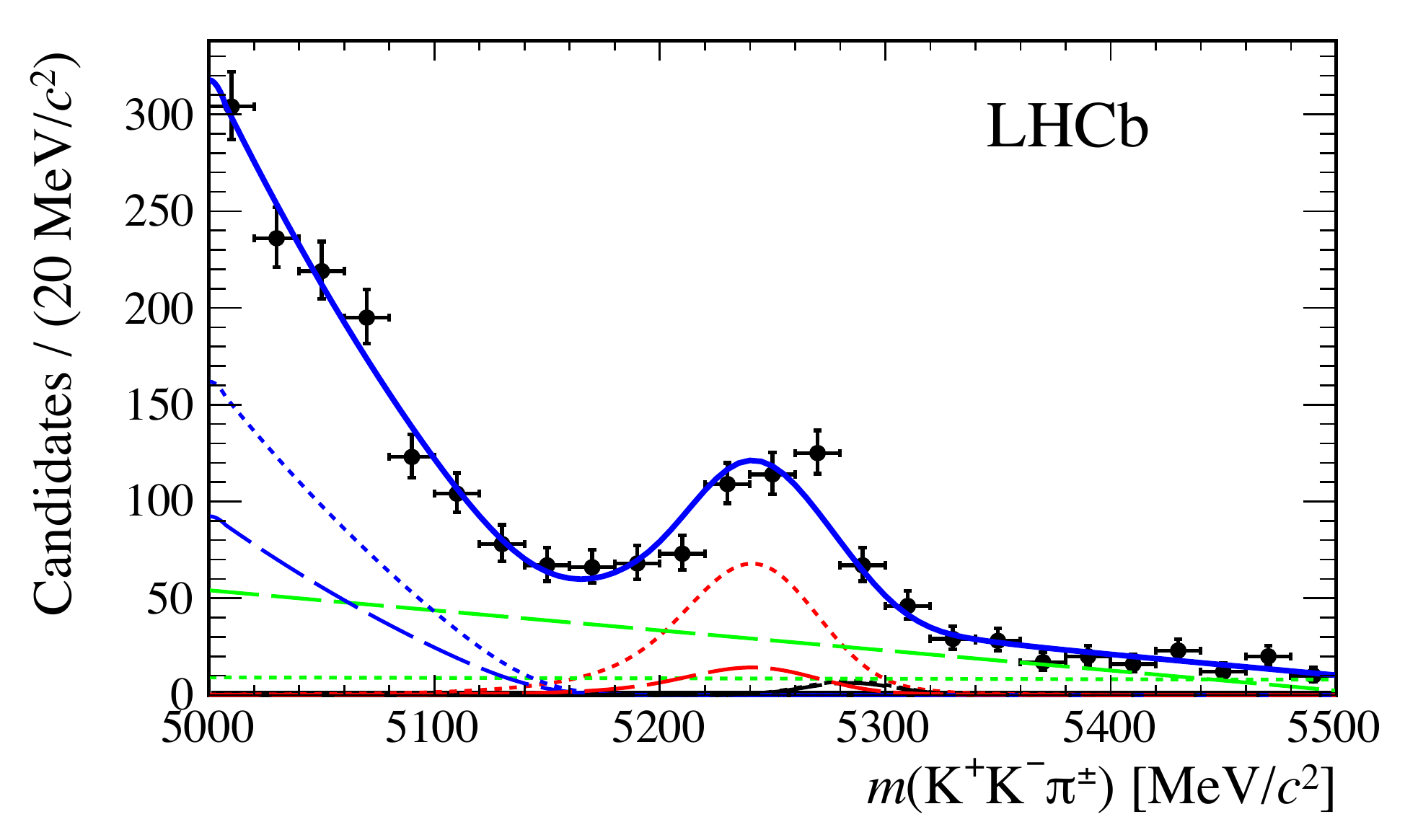}%
\put(85,51){\scriptsize (a)}
\end{overpic}
\begin{overpic}[width=0.48\textwidth]{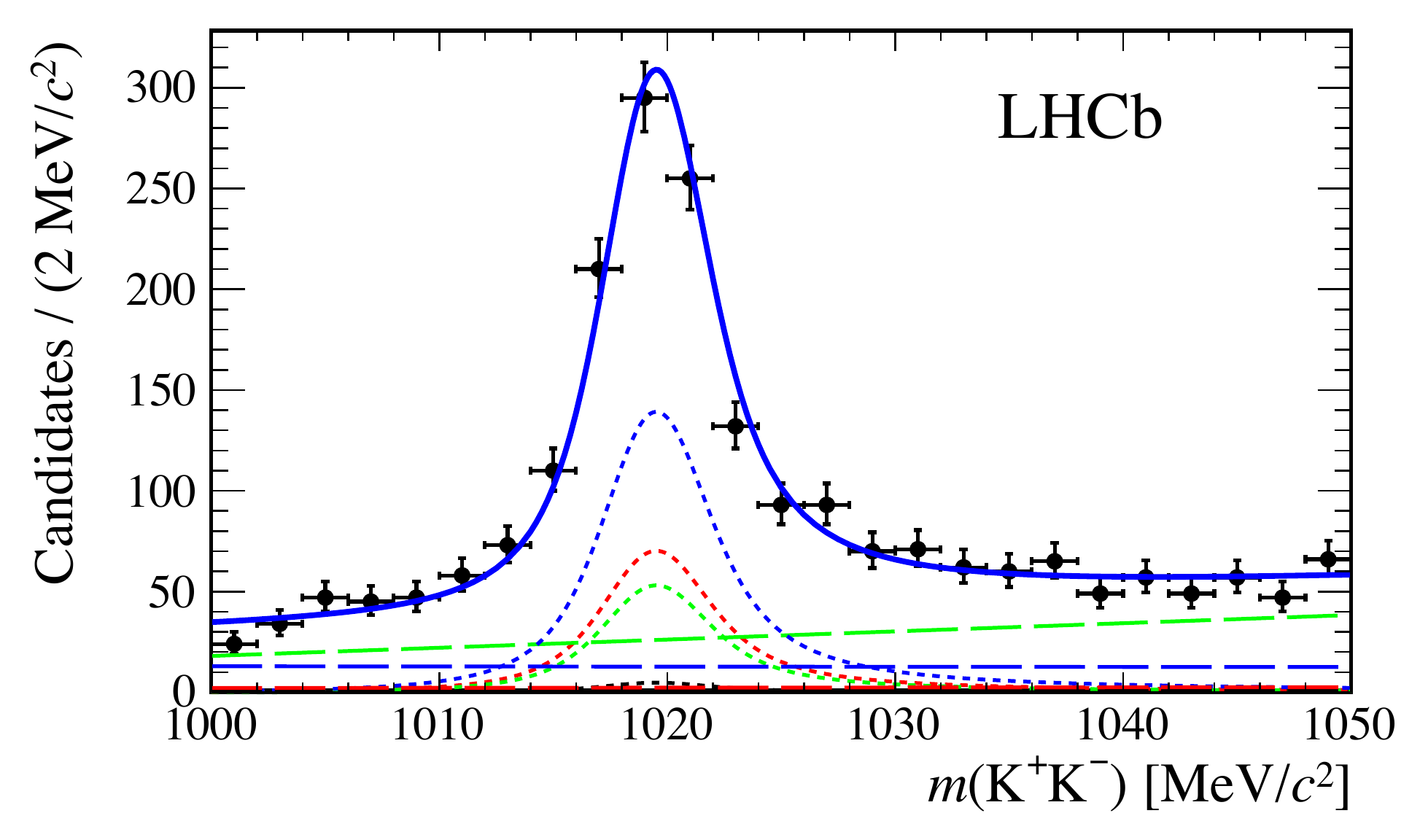}
\put(85,51){\scriptsize (b)}
\end{overpic}
\begin{overpic}[width=0.48\textwidth]{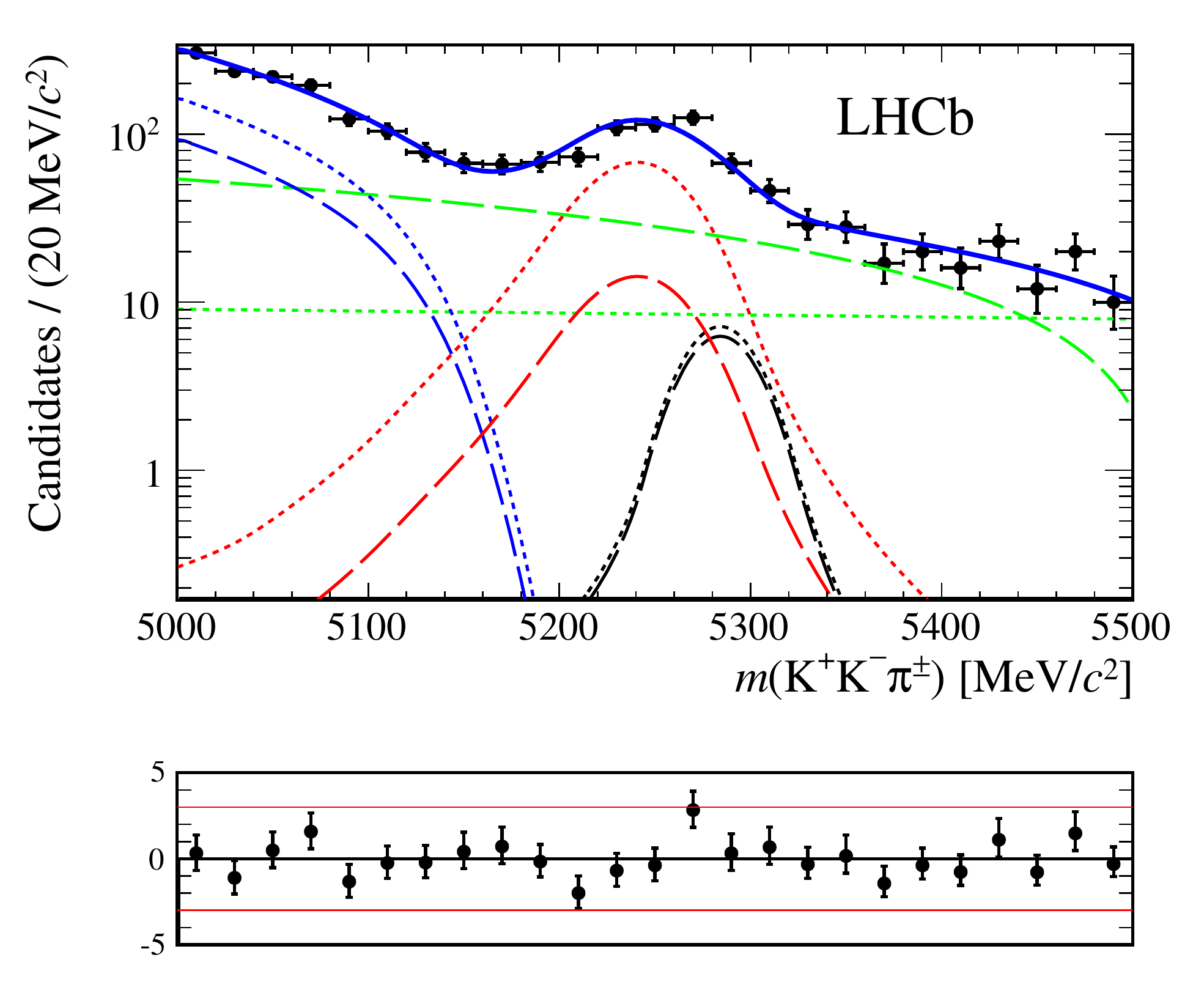}%
\put(85,73){\scriptsize (c)}
\end{overpic}
\begin{overpic}[width=0.48\textwidth]{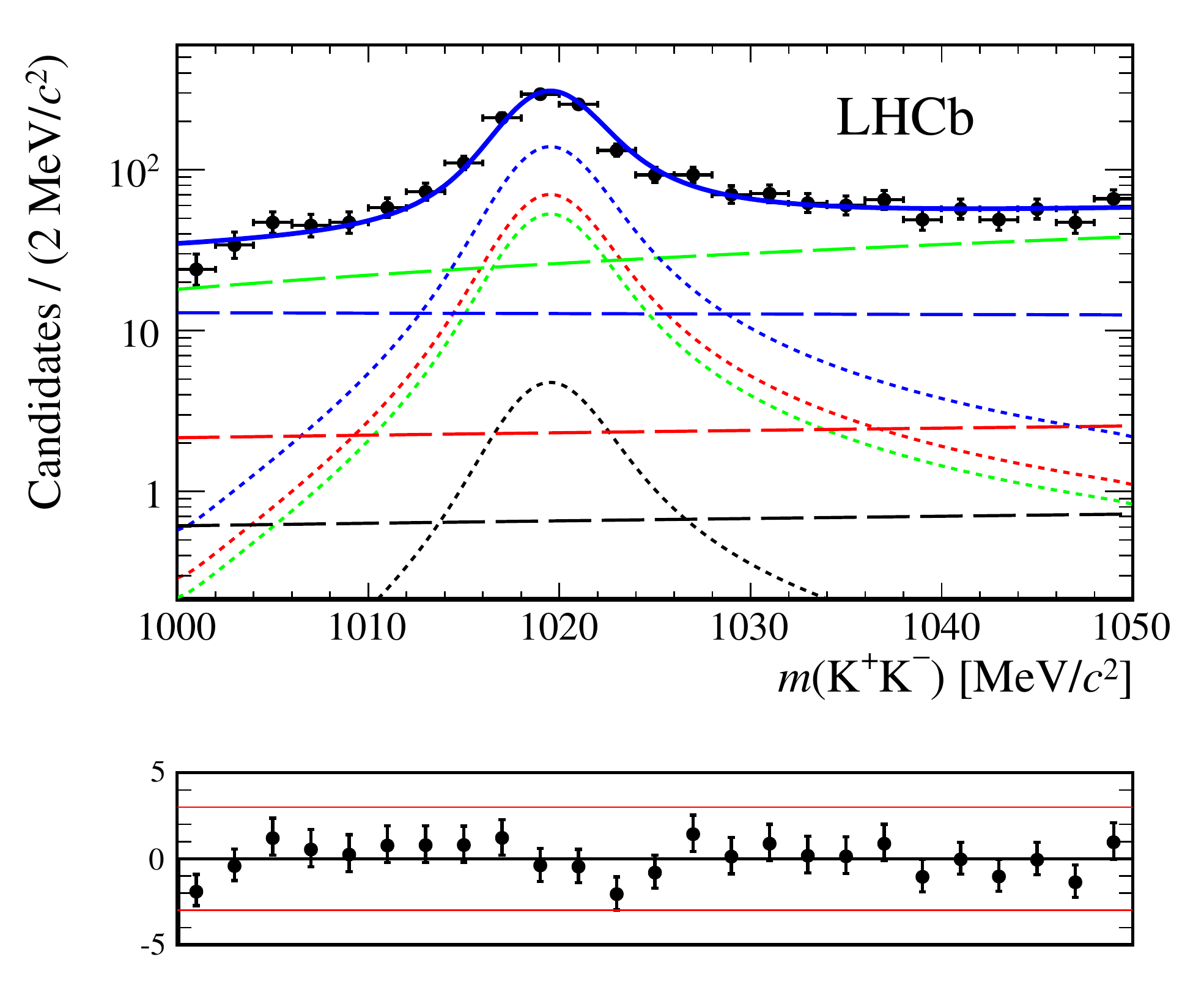}
\put(85,73){\scriptsize (d)}
\end{overpic}
\caption{\small
Distributions of the (a, c) $\Kp\Km\pipm$ and (b, d) $\Kp\Km$ masses
of the selected \decay{\Bpm}{\phi\pipm} candidates, shown both with linear and logarithmic scales.
The solid blue curves represent the result
of the simultaneous fit described in the text, with the following components:
\decay{\Bpm}{\phi\pipm} signal (dotted black), 
nonresonant \decay{\Bpm}{\Kp\Km\pipm} background (dashed black),
\decay{\Bpm}{\phi\Kpm} signal (dotted red), 
nonresonant \decay{\Bpm}{\Kp\Km\Kpm} background (dashed red),
partially-reconstructed $b$-hadron background with (dotted blue) or without 
(dashed blue) a true $\phi$ meson,
and combinatorial background with (dotted green) or without (dashed green) a true $\phi$ meson.
Normalized residuals are displayed below the histograms.}
\label{fig:FinalPhiPi}
\end{figure}

The search for \decay{\Bpm}{\phi\pipm} decays is performed using a simultaneous fit to the \decay{\Bpm}{\phi\pipm} 
\mbox{(${\rm DLL}_{K\pi} < -1$)}
and \decay{\Bpm}{\phi\Kpm}
\mbox{(${\rm DLL}_{K\pi} \ge -1$)}
candidates, 
dividing the \decay{\Bpm}{\phi\pipm} candidates in four subsamples 
according to their $\delta m$ values, each with its set of eight yields. 
The fit has a total of $52$ free parameters: 
$15$ mass shape parameters, $36$ yields, and 
the ratio of the total \decay{\Bpm}{\phi\pipm} yield to the total
\decay{\Bpm}{\phi\Kpm} yield. 

Figure~\ref{fig:FinalPhiPi} 
\afterpage{\clearpage}
shows the projections
of the fitted function superimposed on the observed
mass distributions of the \decay{\Bpm}{\phi\pipm} candidates. 
The total \decay{\Bpm}{\phi\pipm} signal yield is found to be
$19\pm19$, 
while the total \decay{\Bpm}{\phi\Kpm} yield
is $(3486\pm76)+(280\pm25)$ summing the samples of 
\decay{\Bpm}{\phi\Kpm} and \decay{\Bpm}{\phi\pipm} candidates.
The fitted yield ratio is
\begin{equation}
\frac{N(\decay{\Bpm}{\phi\pipm})}{N(\decay{\Bpm}{\phi\Kpm})} = 
(\rm 5.1\,^{+5.3}_{-5.0}(stat)\pm 2.1(syst))\times 10^{-3}\,,
\label{eq:yield_ratio}
\end{equation}
where the systematic uncertainty is the quadratic sum of contributions
due to the modelling of the 
mass shapes ($\pm2.1\times10^{-3}$), 
the fit procedure ($\pm 0.2\times10^{-3}$), 
and interference effects between the
$\phi$ resonance and a $K^+K^-$ pair in an S-wave state
($\pm 0.4\times10^{-3}$). 
The first contribution is obtained by repeating the fit with 
the parameter values of the Crystal Ball and ARGUS functions changed within their uncertainties,
or with an exponential (rather than linear) combinatorial background model. The dominant effect
is due to the 8\% uncertainty on the ratio
$\rho$ of the \decay{\Bpm}{\phi\Kpm} mass resolutions in 
the two ${\rm DLL}_{K\pi}$ regions. Simulation studies show that the 
fit procedure is unbiased, and the statistical precision of this check is assigned as a systematic uncertainty.
%The effect of a possible residual S-wave contribution not fully accounted for by the linear component is 
%investigated by comparing the observed angular distribution of the \decay{\Bpm}{\phi\Kpm} signal 
%with the expectation in absence of any peaking structure in the $K^+K^-$ mass other than that due to a pure 
%P-wave state, and observed differences are propagated as a systematic uncertainty. 

The measurement of the branching fraction ratio is 
obtained as the ratio between Eq.~\ref{eq:yield_ratio} and Eq.~\ref{eq:eff_ratio}:
\begin{equation}
\frac{\BR(\decay{\Bpm}{\phi\pipm})}{\BR(\decay{\Bpm}{\phi\Kpm})} = 
\rm (6.6\,^{+6.9}_{-6.6} (stat) \pm 2.8 (syst))\times 10^{-3} \,.
\label{eq:BR_ratio}
\end{equation}
Since the result is not significantly different from zero,
we also quote upper limits from the integral of the likelihood 
function of this ratio, considering only the physical (non-negative) region. Including systematic uncertainties 
we obtain $\BR(\decay{\Bpm}{\phi\pipm})/\BR(\decay{\Bpm}{\phi\Kpm}) < 0.018\,(0.020)$ at 90\% (95\%) CL. 
Using the current world average
$\BR(\decay{\Bpm}{\phi\Kpm}) = (8.8\,^{+0.7}_{-0.6})\times10^{-6}$~\cite{Beringer:1900zz},
we finally obtain
\begin{eqnarray}
\BR(\decay{\Bpm}{\phi\pipm}) &= & 
(5.8\,^{+6.1}_{-5.8}\pm2.5)\times10^{-8} \\
& < & 1.5\,(1.8)\times10^{-7}~~~\mbox{at 90\% (95\%) CL}\,.
\end{eqnarray}

\section{Conclusions}
\label{sec:Conclusion}

The difference in charge asymmetries between the \decay{\Bpm}{\phi\Kpm} 
and \decay{\Bpm}{\jpsi\Kpm} decay modes is measured in a sample 
of $pp$ collisions at 7\tev centre-of-mass energy, corresponding to an integrated luminosity of 1.0\invfb  collected with the 
LHCb detector. Using the known 
value of the \decay{\Bpm}{\jpsi\Kpm} asymmetry, the \CP-violating charge asymmetry of
\decay{\Bpm}{\phi\Kpm} decays is determined to be
$\ACP(\decay{\Bpm}{\phi \Kpm}) =
\rm 0.022\pm 0.021 (stat) \pm 0.009 (syst)$. This result is almost a factor two 
more precise than the current world average~\cite{Beringer:1900zz}. It is 
consistent with both the absence of \CP violation
and the Standard Model prediction. 

A search for \decay{\Bpm}{\phi\pipm} decays is also performed. 
No significant signal is found.
Using the known branching fraction of the 
\decay{\Bpm}{\phi\Kpm} normalization channel,
an upper limit of $\BR(\decay{\Bpm}{\phi \pipm})< 1.5\,(1.8)\times 10^{-7}$
is set at 90\% (95\%) confidence level. This improves on the previous 
best upper limit~\cite{Aubert:2006nn},
while reaching the upper end of the Standard Model predictions. 
\clearpage

%\input{principles}
%\input{layout}
%\input{detector}
%\input{reference}
%\input{supplementary}

% Do not include this in analysis note and conference reports
\section*{Acknowledgements} 
\noindent We express our gratitude to our colleagues in the CERN
accelerator departments for the excellent performance of the LHC. We
thank the technical and administrative staff at the LHCb
institutes. We acknowledge support from CERN and from the national
agencies: CAPES, CNPq, FAPERJ and FINEP (Brazil); NSFC (China);
CNRS/IN2P3 and Region Auvergne (France); BMBF, DFG, HGF and MPG
(Germany); SFI (Ireland); INFN (Italy); FOM and NWO (The Netherlands);
SCSR (Poland); MEN/IFA (Romania); MinES, Rosatom, RFBR and NRC
``Kurchatov Institute'' (Russia); MinECo, XuntaGal and GENCAT (Spain);
SNSF and SER (Switzerland); NAS Ukraine (Ukraine); STFC (United
Kingdom); NSF (USA). We also acknowledge the support received from the
ERC under FP7. The Tier1 computing centres are supported by IN2P3
(France), KIT and BMBF (Germany), INFN (Italy), NWO and SURF (The
Netherlands), PIC (Spain), GridPP (United Kingdom). We are thankful
for the computing resources put at our disposal by Yandex LLC
(Russia), as well as to the communities behind the multiple open
source software packages that we depend on.

% \input{appendix}

% This should be taken out in the final paper

\addcontentsline{toc}{section}{References}
\setboolean{inbibliography}{true}
\bibliographystyle{LHCb}
%\bibliography{main,LHCb-PAPER,LHCb-CONF,LHCb-DP}
\bibliography{main_single}

\ifx\mcitethebibliography\mciteundefinedmacro
\PackageError{LHCb.bst}{mciteplus.sty has not been loaded}
{This bibstyle requires the use of the mciteplus package.}\fi
\providecommand{\href}[2]{#2}
\begin{mcitethebibliography}{10}
\mciteSetBstSublistMode{n}
\mciteSetBstMaxWidthForm{subitem}{\alph{mcitesubitemcount})}
\mciteSetBstSublistLabelBeginEnd{\mcitemaxwidthsubitemform\space}
{\relax}{\relax}

\bibitem{Beringer:1900zz}
Particle Data Group, J.~Beringer {\em et~al.},
  \ifthenelse{\boolean{articletitles}}{{\it {\href{http://pdg.lbl.gov/}{Review
  of particle physics}}},
  }{}\href{http://dx.doi.org/10.1103/PhysRevD.86.010001}{Phys.\ Rev.\  {\bf
  D86} (2012) 010001}, and 2013 partial update for the 2014 edition\relax
\mciteBstWouldAddEndPuncttrue
\mciteSetBstMidEndSepPunct{\mcitedefaultmidpunct}
{\mcitedefaultendpunct}{\mcitedefaultseppunct}\relax
\EndOfBibitem
\bibitem{Li:2006jv}
H.-n. Li and S.~Mishima, \ifthenelse{\boolean{articletitles}}{{\it
  {Penguin-dominated $B \to P V$ decays in NLO perturbative QCD}},
  }{}\href{http://dx.doi.org/10.1103/PhysRevD.74.094020}{Phys.\ Rev.\  {\bf
  D74} (2006) 094020}, \href{http://arxiv.org/abs/hep-ph/0608277}{{\tt
  arXiv:hep-ph/0608277}}\relax
\mciteBstWouldAddEndPuncttrue
\mciteSetBstMidEndSepPunct{\mcitedefaultmidpunct}
{\mcitedefaultendpunct}{\mcitedefaultseppunct}\relax
\EndOfBibitem
\bibitem{Beneke:2003zv}
M.~Beneke and M.~Neubert, \ifthenelse{\boolean{articletitles}}{{\it {QCD
  factorization for $ B \to PP$ and $B \to PV$ decays}},
  }{}\href{http://dx.doi.org/10.1016/j.nuclphysb.2003.09.026}{Nucl.\ Phys.\
  {\bf B675} (2003) 333}, \href{http://arxiv.org/abs/hep-ph/0308039}{{\tt
  arXiv:hep-ph/0308039}}\relax
\mciteBstWouldAddEndPuncttrue
\mciteSetBstMidEndSepPunct{\mcitedefaultmidpunct}
{\mcitedefaultendpunct}{\mcitedefaultseppunct}\relax
\EndOfBibitem
\bibitem{Lees:2012kxa}
\babar collaboration, J.~P. Lees {\em et~al.},
  \ifthenelse{\boolean{articletitles}}{{\it {Study of \CP violation in
  Dalitz-plot analyses of $\Bd \to \Kp\Km\KS$, $\Bp \to \Kp\Km\Kp$, and $\Bp
  \to \KS\KS\Kp$}},
  }{}\href{http://dx.doi.org/10.1103/PhysRevD.85.112010}{Phys.\ Rev.\  {\bf
  D85} (2012) 112010}, \href{http://arxiv.org/abs/1201.5897}{{\tt
  arXiv:1201.5897}}\relax
\mciteBstWouldAddEndPuncttrue
\mciteSetBstMidEndSepPunct{\mcitedefaultmidpunct}
{\mcitedefaultendpunct}{\mcitedefaultseppunct}\relax
\EndOfBibitem
\bibitem{LHCb-PAPER-2013-027}
LHCb collaboration, R.~Aaij {\em et~al.},
  \ifthenelse{\boolean{articletitles}}{{\it {Measurement of \CP violation in
  the phase space of $B^\pm \to K^\pm\pi^+\pi^-$ and $B^\pm \to K^\pm
  K^+K^-$}}, }{}\href{http://dx.doi.org/10.1103/PhysRevLett.111.101801}{Phys.\
  Rev.\ Lett.\  {\bf 111} (2013) 101801},
  \href{http://arxiv.org/abs/1306.1246}{{\tt arXiv:1306.1246}}\relax
\mciteBstWouldAddEndPuncttrue
\mciteSetBstMidEndSepPunct{\mcitedefaultmidpunct}
{\mcitedefaultendpunct}{\mcitedefaultseppunct}\relax
\EndOfBibitem
\bibitem{Cabibbo:1963yz}
N.~Cabibbo, \ifthenelse{\boolean{articletitles}}{{\it {Unitary symmetry and
  leptonic decays}},
  }{}\href{http://dx.doi.org/10.1103/PhysRevLett.10.531}{Phys.\ Rev.\ Lett.\
  {\bf 10} (1963) 531}\relax
\mciteBstWouldAddEndPuncttrue
\mciteSetBstMidEndSepPunct{\mcitedefaultmidpunct}
{\mcitedefaultendpunct}{\mcitedefaultseppunct}\relax
\EndOfBibitem
\bibitem{Kobayashi:1973fv}
M.~Kobayashi and T.~Maskawa, \ifthenelse{\boolean{articletitles}}{{\it {CP
  violation in the renormalizable theory of weak interaction}},
  }{}\href{http://dx.doi.org/10.1143/PTP.49.652}{Prog.\ Theor.\ Phys.\  {\bf
  49} (1973) 652}\relax
\mciteBstWouldAddEndPuncttrue
\mciteSetBstMidEndSepPunct{\mcitedefaultmidpunct}
{\mcitedefaultendpunct}{\mcitedefaultseppunct}\relax
\EndOfBibitem
\bibitem{Okubo:1963fa}
S.~Okubo, \ifthenelse{\boolean{articletitles}}{{\it {$\phi$-meson and unitary
  symmetry model}},
  }{}\href{http://dx.doi.org/10.1016/S0375-9601(63)92548-9}{Phys.\ Lett.\  {\bf
  5} (1963) 165}\relax
\mciteBstWouldAddEndPuncttrue
\mciteSetBstMidEndSepPunct{\mcitedefaultmidpunct}
{\mcitedefaultendpunct}{\mcitedefaultseppunct}\relax
\EndOfBibitem
\bibitem{Zweig:bothparts}
G.~Zweig, \ifthenelse{\boolean{articletitles}}{{\it {An SU$_3$ model for strong
  interaction symmetry and its breaking}}, }{}\href{\mycds/352337}{CERN-TH-401}
  and \href{\mycds/570209/}{CERN-TH-412} (1964)\relax
\mciteBstWouldAddEndPuncttrue
\mciteSetBstMidEndSepPunct{\mcitedefaultmidpunct}
{\mcitedefaultendpunct}{\mcitedefaultseppunct}\relax
\EndOfBibitem
\bibitem{Iizuka:1966fk}
J.~Iizuka, \ifthenelse{\boolean{articletitles}}{{\it {A systematics and
  phenomenology of meson family}},
  }{}\href{http://dx.doi.org/10.1143/PTPS.37.21}{Prog.\ Theor.\ Phys.\ Suppl.\
  {\bf 37-38} (1966) 21}\relax
\mciteBstWouldAddEndPuncttrue
\mciteSetBstMidEndSepPunct{\mcitedefaultmidpunct}
{\mcitedefaultendpunct}{\mcitedefaultseppunct}\relax
\EndOfBibitem
\bibitem{BarShalom:2002sv}
S.~Bar-Shalom, G.~Eilam, and Y.-D. Yang,
  \ifthenelse{\boolean{articletitles}}{{\it {$B \to \phi \pi$ and $B^0 \to \phi
  \phi$ in the Standard Model and new bounds on R parity violation}},
  }{}\href{http://dx.doi.org/10.1103/PhysRevD.67.014007}{Phys.\ Rev.\  {\bf
  D67} (2003) 014007}, \href{http://arxiv.org/abs/hep-ph/0201244}{{\tt
  arXiv:hep-ph/0201244}}\relax
\mciteBstWouldAddEndPuncttrue
\mciteSetBstMidEndSepPunct{\mcitedefaultmidpunct}
{\mcitedefaultendpunct}{\mcitedefaultseppunct}\relax
\EndOfBibitem
\bibitem{Li:2009zj}
Y.~Li, C.-D. Lu, and W.~Wang, \ifthenelse{\boolean{articletitles}}{{\it
  {Revisiting $B \to \phi \pi$ decays in the Standard Model}},
  }{}\href{http://dx.doi.org/10.1103/PhysRevD.80.014024}{Phys.\ Rev.\  {\bf
  D80} (2009) 014024}, \href{http://arxiv.org/abs/0901.0648}{{\tt
  arXiv:0901.0648}}\relax
\mciteBstWouldAddEndPuncttrue
\mciteSetBstMidEndSepPunct{\mcitedefaultmidpunct}
{\mcitedefaultendpunct}{\mcitedefaultseppunct}\relax
\EndOfBibitem
\bibitem{Kucukarslan:2006wk}
A.~Kucukarslan and U.-G. Meissner, \ifthenelse{\boolean{articletitles}}{{\it
  {$\omega-\phi$ mixing in chiral perturbation theory}},
  }{}\href{http://dx.doi.org/10.1142/S0217732306020743}{Mod.\ Phys.\ Lett.\
  {\bf A21} (2006) 1423}, \href{http://arxiv.org/abs/hep-ph/0603061}{{\tt
  arXiv:hep-ph/0603061}}\relax
\mciteBstWouldAddEndPuncttrue
\mciteSetBstMidEndSepPunct{\mcitedefaultmidpunct}
{\mcitedefaultendpunct}{\mcitedefaultseppunct}\relax
\EndOfBibitem
\bibitem{Benayoun:1999fv}
M.~Benayoun {\em et~al.}, \ifthenelse{\boolean{articletitles}}{{\it {Radiative
  decays, nonet symmetry and SU(3) breaking}},
  }{}\href{http://dx.doi.org/10.1103/PhysRevD.59.114027}{Phys.\ Rev.\  {\bf
  D59} (1999) 114027}, \href{http://arxiv.org/abs/hep-ph/9902326}{{\tt
  arXiv:hep-ph/9902326}}\relax
\mciteBstWouldAddEndPuncttrue
\mciteSetBstMidEndSepPunct{\mcitedefaultmidpunct}
{\mcitedefaultendpunct}{\mcitedefaultseppunct}\relax
\EndOfBibitem
\bibitem{Benayoun:2007cu}
M.~Benayoun {\em et~al.}, \ifthenelse{\boolean{articletitles}}{{\it {The dipion
  mass spectrum in $e^+e^-$ annihilation and $\tau$ decay: a dynamical ($\rho$,
  $\omega$, $\phi$) mixing approach}},
  }{}\href{http://dx.doi.org/10.1140/epjc/s10052-008-0586-6}{Eur.\ Phys.\ J.\
  {\bf C55} (2008) 199}, \href{http://arxiv.org/abs/0711.4482}{{\tt
  arXiv:0711.4482}}\relax
\mciteBstWouldAddEndPuncttrue
\mciteSetBstMidEndSepPunct{\mcitedefaultmidpunct}
{\mcitedefaultendpunct}{\mcitedefaultseppunct}\relax
\EndOfBibitem
\bibitem{Benayoun:2009im}
M.~Benayoun, P.~David, L.~DelBuono, and O.~Leitner,
  \ifthenelse{\boolean{articletitles}}{{\it {A global treatment of VMD physics
  up to the $\phi$: I. $e^+$ $e^-$ annihilations, anomalies and vector meson
  partial widths}},
  }{}\href{http://dx.doi.org/10.1140/epjc/s10052-009-1197-6}{Eur.\ Phys.\ J.\
  {\bf C65} (2010) 211}, \href{http://arxiv.org/abs/0907.4047}{{\tt
  arXiv:0907.4047}}\relax
\mciteBstWouldAddEndPuncttrue
\mciteSetBstMidEndSepPunct{\mcitedefaultmidpunct}
{\mcitedefaultendpunct}{\mcitedefaultseppunct}\relax
\EndOfBibitem
\bibitem{Gronau:2008kk}
M.~Gronau and J.~L. Rosner, \ifthenelse{\boolean{articletitles}}{{\it {$B$
  decays dominated by $\omega-\phi$ mixing}},
  }{}\href{http://dx.doi.org/10.1016/j.physletb.2008.07.016}{Phys.\ Lett.\
  {\bf B666} (2008) 185}, \href{http://arxiv.org/abs/0806.3584}{{\tt
  arXiv:0806.3584}}\relax
\mciteBstWouldAddEndPuncttrue
\mciteSetBstMidEndSepPunct{\mcitedefaultmidpunct}
{\mcitedefaultendpunct}{\mcitedefaultseppunct}\relax
\EndOfBibitem
\bibitem{Aaij:2013mtm}
LHCb collaboration, R.~Aaij {\em et~al.},
  \ifthenelse{\boolean{articletitles}}{{\it {First observation of
  $\decay{\Bd}{\jpsi\Kp\Km}$ and search for $\decay{\Bd}{\jpsi\phi}$ decays}},
  }{}\href{http://arxiv.org/abs/1308.5916}{{\tt arXiv:1308.5916}}\relax
\mciteBstWouldAddEndPuncttrue
\mciteSetBstMidEndSepPunct{\mcitedefaultmidpunct}
{\mcitedefaultendpunct}{\mcitedefaultseppunct}\relax
\EndOfBibitem
\bibitem{Aubert:2006nn}
BaBar collaboration, B.~Aubert {\em et~al.},
  \ifthenelse{\boolean{articletitles}}{{\it {Search for $B^{+} \to \phi
  \pi^{+}$ and $B^0 \to \phi \pi^0$ decays}},
  }{}\href{http://dx.doi.org/10.1103/PhysRevD.74.011102}{Phys.\ Rev.\  {\bf
  D74} (2006) 011102}, \href{http://arxiv.org/abs/hep-ex/0605037}{{\tt
  arXiv:hep-ex/0605037}}\relax
\mciteBstWouldAddEndPuncttrue
\mciteSetBstMidEndSepPunct{\mcitedefaultmidpunct}
{\mcitedefaultendpunct}{\mcitedefaultseppunct}\relax
\EndOfBibitem
\bibitem{Abazov:2013sqa}
D0 collaboration, V.~M. Abazov {\em et~al.},
  \ifthenelse{\boolean{articletitles}}{{\it {Measurement of direct CP violation
  parameters in \decay{\Bpm}{\jpsi\Kpm} and \decay{\Bpm}{\jpsi\pipm} decays
  with 10.4\invfb of Tevatron data}},
  }{}\href{http://dx.doi.org/10.1103/PhysRevLett.110.241801}{Phys.\ Rev.\
  Lett.\  {\bf 110} (2013) 241801}, \href{http://arxiv.org/abs/1304.1655}{{\tt
  arXiv:1304.1655}}\relax
\mciteBstWouldAddEndPuncttrue
\mciteSetBstMidEndSepPunct{\mcitedefaultmidpunct}
{\mcitedefaultendpunct}{\mcitedefaultseppunct}\relax
\EndOfBibitem
\bibitem{Alves:2008zz}
LHCb collaboration, A.~A. Alves~Jr. {\em et~al.},
  \ifthenelse{\boolean{articletitles}}{{\it {The \lhcb detector at the LHC}},
  }{}\href{http://dx.doi.org/10.1088/1748-0221/3/08/S08005}{JINST {\bf 3}
  (2008) S08005}\relax
\mciteBstWouldAddEndPuncttrue
\mciteSetBstMidEndSepPunct{\mcitedefaultmidpunct}
{\mcitedefaultendpunct}{\mcitedefaultseppunct}\relax
\EndOfBibitem
\bibitem{LHCb-DP-2012-003}
M.~Adinolfi {\em et~al.}, \ifthenelse{\boolean{articletitles}}{{\it
  {Performance of the \lhcb RICH detector at the LHC}},
  }{}\href{http://dx.doi.org/10.1140/epjc/s10052-013-2431-9}{Eur.\ Phys.\ J.\
  {\bf C73} (2013) 2431}, \href{http://arxiv.org/abs/1211.6759}{{\tt
  arXiv:1211.6759}}\relax
\mciteBstWouldAddEndPuncttrue
\mciteSetBstMidEndSepPunct{\mcitedefaultmidpunct}
{\mcitedefaultendpunct}{\mcitedefaultseppunct}\relax
\EndOfBibitem
\bibitem{LHCb-DP-2012-004}
R.~Aaij {\em et~al.}, \ifthenelse{\boolean{articletitles}}{{\it {The \lhcb
  trigger and its performance in 2011}},
  }{}\href{http://dx.doi.org/10.1088/1748-0221/8/04/P04022}{JINST {\bf 8}
  (2013) P04022}, \href{http://arxiv.org/abs/1211.3055}{{\tt
  arXiv:1211.3055}}\relax
\mciteBstWouldAddEndPuncttrue
\mciteSetBstMidEndSepPunct{\mcitedefaultmidpunct}
{\mcitedefaultendpunct}{\mcitedefaultseppunct}\relax
\EndOfBibitem
\bibitem{BBDT}
V.~V. Gligorov and M.~Williams, \ifthenelse{\boolean{articletitles}}{{\it
  {Efficient, reliable and fast high-level triggering using a bonsai boosted
  decision tree}},
  }{}\href{http://dx.doi.org/10.1088/1748-0221/8/02/P02013}{JINST {\bf 8}
  (2013) P02013}, \href{http://arxiv.org/abs/1210.6861}{{\tt
  arXiv:1210.6861}}\relax
\mciteBstWouldAddEndPuncttrue
\mciteSetBstMidEndSepPunct{\mcitedefaultmidpunct}
{\mcitedefaultendpunct}{\mcitedefaultseppunct}\relax
\EndOfBibitem
\bibitem{Sjostrand:2006za}
T.~Sj\"{o}strand, S.~Mrenna, and P.~Skands,
  \ifthenelse{\boolean{articletitles}}{{\it {PYTHIA 6.4 physics and manual}},
  }{}\href{http://dx.doi.org/10.1088/1126-6708/2006/05/026}{JHEP {\bf 05}
  (2006) 026}, \href{http://arxiv.org/abs/hep-ph/0603175}{{\tt
  arXiv:hep-ph/0603175}}\relax
\mciteBstWouldAddEndPuncttrue
\mciteSetBstMidEndSepPunct{\mcitedefaultmidpunct}
{\mcitedefaultendpunct}{\mcitedefaultseppunct}\relax
\EndOfBibitem
\bibitem{LHCb-PROC-2010-056}
I.~Belyaev {\em et~al.}, \ifthenelse{\boolean{articletitles}}{{\it {Handling of
  the generation of primary events in \gauss, the \lhcb simulation framework}},
  }{}\href{http://dx.doi.org/10.1109/NSSMIC.2010.5873949}{Nuclear Science
  Symposium Conference Record (NSS/MIC) {\bf IEEE} (2010) 1155}\relax
\mciteBstWouldAddEndPuncttrue
\mciteSetBstMidEndSepPunct{\mcitedefaultmidpunct}
{\mcitedefaultendpunct}{\mcitedefaultseppunct}\relax
\EndOfBibitem
\bibitem{Lange:2001uf}
D.~J. Lange, \ifthenelse{\boolean{articletitles}}{{\it {The EvtGen particle
  decay simulation package}},
  }{}\href{http://dx.doi.org/10.1016/S0168-9002(01)00089-4}{Nucl.\ Instrum.\
  Meth.\  {\bf A462} (2001) 152}\relax
\mciteBstWouldAddEndPuncttrue
\mciteSetBstMidEndSepPunct{\mcitedefaultmidpunct}
{\mcitedefaultendpunct}{\mcitedefaultseppunct}\relax
\EndOfBibitem
\bibitem{Golonka:2005pn}
P.~Golonka and Z.~Was, \ifthenelse{\boolean{articletitles}}{{\it {PHOTOS Monte
  Carlo: a precision tool for QED corrections in $Z$ and $W$ decays}},
  }{}\href{http://dx.doi.org/10.1140/epjc/s2005-02396-4}{Eur.\ Phys.\ J.\  {\bf
  C45} (2006) 97}, \href{http://arxiv.org/abs/hep-ph/0506026}{{\tt
  arXiv:hep-ph/0506026}}\relax
\mciteBstWouldAddEndPuncttrue
\mciteSetBstMidEndSepPunct{\mcitedefaultmidpunct}
{\mcitedefaultendpunct}{\mcitedefaultseppunct}\relax
\EndOfBibitem
\bibitem{Allison:2006ve}
Geant4 collaboration, J.~Allison {\em et~al.},
  \ifthenelse{\boolean{articletitles}}{{\it {Geant4 developments and
  applications}}, }{}\href{http://dx.doi.org/10.1109/TNS.2006.869826}{IEEE
  Trans.\ Nucl.\ Sci.\  {\bf 53} (2006) 270}\relax
\mciteBstWouldAddEndPuncttrue
\mciteSetBstMidEndSepPunct{\mcitedefaultmidpunct}
{\mcitedefaultendpunct}{\mcitedefaultseppunct}\relax
\EndOfBibitem
\bibitem{Agostinelli:2002hh}
Geant4 collaboration, S.~Agostinelli {\em et~al.},
  \ifthenelse{\boolean{articletitles}}{{\it {Geant4: a simulation toolkit}},
  }{}\href{http://dx.doi.org/10.1016/S0168-9002(03)01368-8}{Nucl.\ Instrum.\
  Meth.\  {\bf A506} (2003) 250}\relax
\mciteBstWouldAddEndPuncttrue
\mciteSetBstMidEndSepPunct{\mcitedefaultmidpunct}
{\mcitedefaultendpunct}{\mcitedefaultseppunct}\relax
\EndOfBibitem
\bibitem{LHCb-PROC-2011-006}
M.~Clemencic {\em et~al.}, \ifthenelse{\boolean{articletitles}}{{\it {The \lhcb
  simulation application, \gauss: design, evolution and experience}},
  }{}\href{http://dx.doi.org/10.1088/1742-6596/331/3/032023}{{J.\ Phys.\ Conf.\
  Ser.\ } {\bf 331} (2011) 032023}\relax
\mciteBstWouldAddEndPuncttrue
\mciteSetBstMidEndSepPunct{\mcitedefaultmidpunct}
{\mcitedefaultendpunct}{\mcitedefaultseppunct}\relax
\EndOfBibitem
\bibitem{Skwarnicki:1986xj}
T.~Skwarnicki, {\em {A study of the radiative cascade transitions between the
  Upsilon-prime and Upsilon resonances}}, PhD thesis, Institute of Nuclear
  Physics, Krakow, 1986,
  {\href{http://inspirehep.net/record/230779/files/230779.pdf}{DESY-F31-86-02}}\relax
\mciteBstWouldAddEndPuncttrue
\mciteSetBstMidEndSepPunct{\mcitedefaultmidpunct}
{\mcitedefaultendpunct}{\mcitedefaultseppunct}\relax
\EndOfBibitem
\bibitem{Argus}
ARGUS collaboration, H.~Albrecht {\em et~al.},
  \ifthenelse{\boolean{articletitles}}{{\it {Search for $b\to s\gamma$ in
  exclusive decays of $B$ mesons}},
  }{}\href{http://dx.doi.org/10.1016/0370-2693(89)91177-5}{Phys.\ Lett.\  {\bf
  B229} (1989) 304}\relax
\mciteBstWouldAddEndPuncttrue
\mciteSetBstMidEndSepPunct{\mcitedefaultmidpunct}
{\mcitedefaultendpunct}{\mcitedefaultseppunct}\relax
\EndOfBibitem
\bibitem{Punzi:2004wh}
G.~Punzi, \ifthenelse{\boolean{articletitles}}{{\it {Comments on likelihood
  fits with variable resolution}}, }{}eConf {\bf C030908} (2003) WELT002,
  \href{http://arxiv.org/abs/physics/0401045}{{\tt
  arXiv:physics/0401045}}\relax
\mciteBstWouldAddEndPuncttrue
\mciteSetBstMidEndSepPunct{\mcitedefaultmidpunct}
{\mcitedefaultendpunct}{\mcitedefaultseppunct}\relax
\EndOfBibitem
\end{mcitethebibliography}

\end{document}